\documentclass[
superscriptaddress,
amsmath,
amssymb,
aps,
prl,
twocolumn,
longbibliography
]{revtex4-2}

\usepackage[T1]{fontenc}
\usepackage[utf8]{inputenc}

\usepackage{times} 
\usepackage{xr-hyper}
\usepackage[unicode=true,breaklinks=true,colorlinks=true]{hyperref}
\hypersetup{citecolor=blue,urlcolor=blue,linkcolor=black,filecolor=black}
\usepackage{graphicx}
\usepackage{bm}
\usepackage{bbold}
\usepackage{braket}
\usepackage[dvipsnames]{xcolor}

\usepackage{mathtools}
\usepackage{physics}
\usepackage{tensor}
\definecolor{ballblue}{rgb}{0.13, 0.67, 0.8}
\definecolor{ao}{rgb}{0.0, 0.5, 0.0}

\makeatletter
\newcommand*{\addFileDependency}[1]{
\typeout{(#1)}
\@addtofilelist{#1}
\IfFileExists{#1}{}{\typeout{No file #1.}}
}\makeatother

\usepackage{pgffor}
\usepackage{pdfpages}
\makeatletter
\AtBeginDocument{\let\LS@rot\@undefined}
\makeatother

\usepackage[normalem]{ulem}

\begin{document}
\title{Engineering artificial atomic systems of giant electric dipole moment}
\author{Baiyi Yu}
\author{Yaoming Chu}
\email{yaomingchu@hust.edu.cn}
\affiliation{School of Physics, International Joint Laboratory on Quantum Sensing and Quantum Metrology, Hubei Key Laboratory of Gravitation and Quantum Physics, Institute for Quantum Science and Engineering, Wuhan National High Magnetic Field Center, Huazhong University of Science and Technology, Wuhan 430074, China}
\author{Ralf Betzholz}
\author{Shaoliang Zhang}
\email{shaoliang@hust.edu.cn}
\affiliation{School of Physics, International Joint Laboratory on Quantum Sensing and Quantum Metrology, Hubei Key Laboratory of Gravitation and Quantum Physics, Institute for Quantum Science and Engineering, Wuhan National High Magnetic Field Center, Huazhong University of Science and Technology, Wuhan 430074, China}
\author{Jianming Cai}
\email{jianmingcai@hust.edu.cn}
\affiliation{School of Physics, International Joint Laboratory on Quantum Sensing and Quantum Metrology, Hubei Key Laboratory of Gravitation and Quantum Physics, Institute for Quantum Science and Engineering, Wuhan National High Magnetic Field Center, Huazhong University of Science and Technology, Wuhan 430074, China}
\affiliation{Shanghai Key Laboratory of Magnetic Resonance, East China Normal University, Shanghai 200062, China}

\date{\today}

\begin{abstract}
The electric dipole moment (EDM) plays a crucial role in determining the interaction strength of an atom with electric fields, making it paramount to quantum technologies based on coherent atomic control. We propose a scheme for engineering the potential in a Paul trap to realize a two-level quantum system with a giant EDM formed by the motional states of a trapped electron. We show that, under realistic experimental conditions, the EDM can significantly exceed the ones attainable with Rydberg atoms. Furthermore, we show that such artificial atomic dipoles can be efficiently initialized, readout, and coherently controlled, thereby providing a potential platform for quantum technologies such as ultrahigh-sensitivity electric-field sensing. 
\end{abstract}

\maketitle

{\it Introduction.---} Coherent coupling between atoms and electric fields is one of the most essential ingredients in light-matter interactions. 
Its strength critically depends on the magnitude of the electric dipole moment (EDM) of the atomic system~\cite{scullyQuantumOptics1997,wallsQuantumOptics2008}. 
A large EDM, and thereby a strong coupling, significantly enhances the speed of coherent manipulation~\cite{saffmanQuantumInformationRydberg2010}, 
enables novel driving or coupling regimes~\cite{oliverMachZehnderInterferometryStrongly2005,silveriProbeSpectroscopyQuasienergy2013,barfussStrongMechanicalDriving2015,jimenez-garciaTunableSpinOrbitCoupling2015,raoNonlinearDynamicsTwolevel2017,yanEffectsCounterrotatingCouplings2017,yangFloquetEngineeringEntanglement2019,bruneQuantumRabiOscillation1996,hoganDrivingRydbergRydbergTransitions2012,morganCouplingRydbergAtoms2020,forn-diazUltrastrongCouplingRegimes2019,friskkockumUltrastrongCouplingLight2019,tokuraCoherentSingleElectron2006a,cottetSpinQuantumBit2010,peterssonCircuitQuantumElectrodynamics2012,freyDipoleCouplingDouble2012,kawakamiElectricalControlLonglived2014,viennotCoherentCouplingSingle2015,beaudoinCouplingSingleElectron2016,miStrongCouplingSingle2017,landigCoherentSpinPhoton2018,miCoherentSpinPhotonInterface2018,samkharadzeStrongSpinphotonCoupling2018,scarlinoSituTuningElectricDipole2022}, 
and increases the sensitivity to electric fields~\cite{gleyzesQuantumJumpsLight2007,harocheNobelLectureControlling2013,degenQuantumSensing2017,zhangSensitiveDetectionMillimeter2023}. 
A well-known example of quantum systems with a large EDM are Rydberg atoms~\cite{gallagherRydbergAtoms1994}.
In $^{87}$Rb atoms, for instance, the EDM between neighboring states with principle quantum number $n\sim 65$ is roughly $4000\,ea_0$, with the elementary charge $e$ and Bohr radius $a_0$, which corresponds to $0.2$~$e\mu$m~\cite{sedlacekMicrowaveElectrometryRydberg2012a}. 
This magnitude of the EDM endows Rydberg atoms with exceptional sensitivity to electric fields~\cite{sedlacekMicrowaveElectrometryRydberg2012a,osterwalderUsingHighRydberg1999,faconSensitiveElectrometerBased2016,jingAtomicSuperheterodyneReceiver2020a,hollowayAtomBasedRFElectric2017,liuHighlySensitiveMeasurement2022a,brownVHFUHFDetection2022,rotunnoPseudoresonantDetectionLow2022} and the resulting strong inter-atomic dipole-dipole interaction shows great promise for applications in quantum information processing~\cite{lukinDipoleBlockadeQuantum2001,beugnonTwodimensionalTransportTransfer2007,comparatDipoleBlockadeCold2010,bernienProbingManybodyDynamics2017,keeslingQuantumKibbleZurek2019,browaeysManybodyPhysicsIndividually2020,ebadiQuantumPhasesMatter2021,semeghiniProbingTopologicalSpin2021,everedHighfidelityParallelEntangling2023}.

However, in Rydberg atoms, further augmenting the EDM magnitude by increasing the principle quantum number inevitably results in small binding energies ($\propto n^{-2}$)~\cite{gallagherRydbergAtoms1994,comparatDipoleBlockadeCold2010} and thereby instability of the Rydberg states. This would especially be the case if the transition frequency were to reach the MHz range, a range that is indispensable in broadcasting and air-to-ground communication, owing to the long wavelengths and extended propagation distances~\cite{hollowayAtomBasedRFElectric2017,liuHighlySensitiveMeasurement2022a,brownVHFUHFDetection2022,rotunnoPseudoresonantDetectionLow2022}. 
Therefore, it would be appealing to realize stable quantum systems with giant EDM, even larger than those of Rydberg atoms, particularly in the MHz resonance-frequency range.

Our idea to reach this goal is to create an artificial atom with Rydberg-like states by confining a single electron within a specifically engineered potential. Under the dipole approximation~\cite{scullyQuantumOptics1997,wallsQuantumOptics2008}, the coupling between this artificial atom and electric fields is still governed by the EDM even though there is no ion core. To obtain Rydberg-like states, the engineered potential should bear the essential feature of the Coulomb potential in natural atoms, i.e., the inverse-distance form.
More importantly, it should ensure that the eigenstates entailing a giant EDM are more stable than Rydberg states with high principle quantum numbers. 
The key ingredients to overcome these challenges are a delicate design of such a potential as well as a coherent control of the trapped electron. 

In this Letter, we systematically engineer the trapping potential to obtain a two-level quantum system, formed by motional states of the electron, endowed with a resonance frequency within the MHz range and an EDM magnitude of several $e\mu$m. 
The system can be initialized via fast quasiadiabatic dynamics by appropriately deforming the potential~\cite{martinez-garaotVibrationalModeMultiplexing2013,martinez-garaotFastBiasInversion2015,martinez-garaotFastQuasiadiabaticDynamics2015,abahQuantumStateEngineering2020}. 
To read out the quantum state, we encode the information on the motional degree of freedom in the spin states using a magnetic-field gradient and then perform a projective measurement on the spin degree of freedom~\cite{pengSpinReadoutTrapped2017}. 
Under realistic experimental conditions, our analysis demonstrates that the magnitude of the EDM can reach $7$~$e\mu$m, which is more than an order of magnitude larger than those between stable Rydberg states, with initialization and readout fidelities above $95\%$. We demonstrate that recent progress in trapping and controlling electrons in Paul traps~\cite{pengSpinReadoutTrapped2017,matthiesenTrappingElectronsRoomTemperature2021,yuFeasibilityStudyQuantum2022,sutherlandOneTwoqubitGate2022a} suggests the feasibility of the scheme we present and that the system would provide a superior performance in electric-field sensing.

\begin{figure}{t}
\centering
\includegraphics{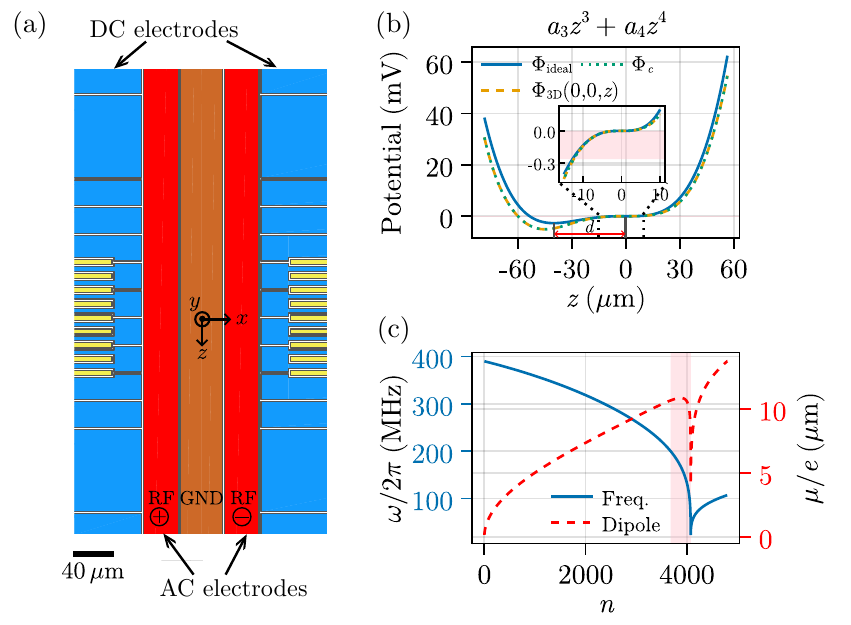}
\caption{\label{fig1}(a) Cropped geometry of the electrodes in our trap design (see SM Sec.~\ref{supp-subsec:electrode geometry} for a more detailed geometry~\cite{supp}).
The AC electrodes (red) are driven by the same RF voltage but with opposite phase, whereas the DC electrodes (blue and yellow) are supplied with voltages symmetric around the $z$ axis. (b) $\Phi_{\mathrm{ideal}}$ is the ideal potential of the form $a_3z^3+a_4z^4$,  $\Phi_{\mathrm{3D}}$ is the actual potential generated by the DC-electrode design, and $\Phi_c$ is the effective axial potential obtained from Eq.~\eqref{eq:effective-potential}. $\Phi_{\mathrm{3D}}$ and $\Phi_c$ are both shifted by a constant, allowing a clearer comparison between the three potentials. The inset shows the range $-15$~$\mu$m $<z<10$~$\mu$m, the pink region of which covers the potential energies mapped from the eigenenergies corresponding to the pink region of (c). (c) Transition frequency (blue) and the corresponding EDM magnitude (red) between the eigenstates $\ket{\psi_{n+1}}$ and $\ket{\psi_n}$ for the effective axial potential $\Phi_c$ shown in (b). The pink region covers quantum numbers with $3679\leq n\leq 4079$. For (b) and (c), we have used the parameter $d=40$~$\mu$m.}
\end{figure}

{\it Anharmonic potential engineering.---} Our goal is to construct a stable two-level system with a giant EDM by designing a suitable trap potential. 
The key element in our design is to extend the potential generated by the DC electrodes [blue and yellow in Fig.~\hyperref[fig1]{\ref{fig1}(a)}] from merely second-order, as in usual Paul traps, to third- and fourth-order in the axial coordinate $z$, see Fig.~\hyperref[fig1]{\ref{fig1}(b)}. This results in a potential similar to the attractive Coulomb potential over a certain range of $z$. We find that, for such a potential, there are large-quantum-number eigenstates with highly nonlinear energies and giant EDMs [cf. Fig.~\hyperref[fig1]{\ref{fig1}(c)}], where the EDM between the $i$th and $j$th eigenstates, $\ket{\psi_i}$ and $\ket{\psi_j}$, of the motion in the $z$ direction is defined as
\begin{equation}
    \mu_{ij}=e\mel{\psi_i}{z}{\psi_j}.
\end{equation}
With the nonlinearity of eigenenergies, these neighboring eigenstates with giant EDM can strongly interact with the resonant electric field as two-level systems.
Moreover, in contrast to Rydberg states, these large-quantum-number eigenstates are still stably trapped in our designed trap potential.

Figure~\hyperref[fig1]{\ref{fig1}(a)} shows the prototype of our trap, which combines two symmetric layers of electrodes separated along the $y$ direction. 
As in usual Paul traps~\cite{leibfriedQuantumDynamicsSingle2003,paulElectromagneticTrapsCharged1990,winelandExperimentalIssuesCoherent1998,matthiesenTrappingElectronsRoomTemperature2021,yuFeasibilityStudyQuantum2022,sutherlandOneTwoqubitGate2022a}, the AC electrodes [red in Fig.~\hyperref[fig1]{\ref{fig1}(a)}] are driven by a radio-frequency (RF) voltage, generating an effective confinement in the radial ($x$ and $y$) directions with a secular frequency $\omega_r$ (see Sec.~\ref{supp-subsec:RF-trapping} of the Supplemental Material (SM)~\cite{supp}). 
In Fig.~\hyperref[fig1]{\ref{fig1}(b)}, the yellow line shows the actual potential $\Phi_{\mathrm{3D}}$ generated by the DC electrodes, which ideally has the form $a_3 z^3+a_4 z^4$ along $x=y=0$, as shown by the blue line. The coefficients $a_3$ and $a_4$ can be shown to have the form (see Sec.~\ref{supp-subsec:a3a4} of the SM~\cite{supp})
\begin{equation}\label{eq:a3a4}
a_3=\frac{2a_2^\prime}{3d},\quad a_4=\frac{3a_3}{4d},
\end{equation}
where $d$ represents the distance between the two points satisfying $\pdv*{\Phi_{\mathrm{3D}}}{z}=0$ along $x=y=0$ and $a_2^\prime=m_e\omega_h^{\prime\,2}/2e$ describes the approximately harmonic confinement in the $z$ direction, centered around $z=-d$, with frequency $\omega_h^{\prime}$. Without loss of generality, here, we choose the parameters $d=40$~$\mu$m and $\omega_h^{\prime}=(2\pi)~300$~MHz for numerical demonstrations. 

Figure~\hyperref[fig1]{\ref{fig1}(c)} is obtained with the eigenenergies and eigenstates of the axial-motion Hamiltonian $H_z=\frac{p_z^2}{2m_e}+e\Phi_c(z)$, where $\Phi_c(z)$ is the effective axial potential 
\begin{equation}\label{eq:effective-potential}
    \Phi_{c}(z)=\iint \qty(\Phi_{\mathrm{3D}}+\Phi_\mathrm{pp})\abs{\psi(x)}^2\abs{\psi(y)}^2 \,dx \,dy,
\end{equation}
with $\psi(x)$ and $\psi(y)$ representing the electron wave function in the $x$ and $y$ direction, respectively, and $\Phi_\mathrm{pp}$ denoting the static RF pseudopotential. 
Equation~\eqref{eq:effective-potential} accounts for axial-radial motion coupling induced by the third- and fourth-order DC-potential terms, and for imperfections of both the DC and RF potentials arising from realistic experimental conditions, see SM Sec.~\ref{supp-subsec:effective potential}~\cite{supp}.
The blue line in Fig.~\hyperref[fig1]{\ref{fig1}(c)} shows the transition frequency between eigenstates $\ket{\psi_{n+1}}$ and $\ket{\psi_n}$. 
The red line shows the corresponding EDM magnitude, which easily exceeds $10$~$e\mu$m within the pink region.

\begin{figure}[t]
\centering
\includegraphics{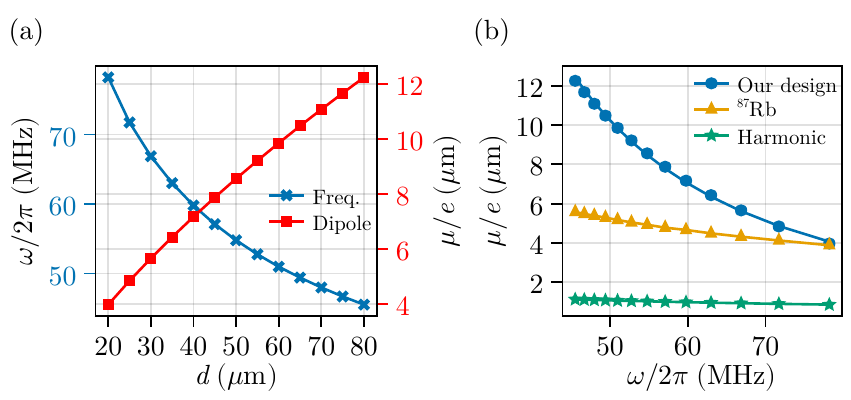}
\caption{
        (a) Dependence of the transition frequency and the EDM magnitude on the potential parameter $d$. (b) EDM magnitude as a function of the transition frequency. Blue: Two-level systems resulting from the potential designed in (a). Yellow: Rydberg states $(n+1) P_{3/2}m_j{=}3/2$ and $n D_{5/2}m_j{=}5/2$ of $^{87}$Rb with $n$ in the range 294--352 (calculated using the Alkali Rydberg Calculator package~\cite{sibalicARCOpensourceLibrary2017}). Green: Harmonic-oscillator ground state and single-phonon Fock state.
    }
    \label{fig2}
\end{figure}

 In Fig.~\hyperref[fig2]{\ref{fig2}(a)}, we show that the magnitude of the EDM increases for a larger $d$, whereas the transition frequency decreases (see SM Sec.~\ref{supp-sec:increase d}~\cite{supp}). 
 A comparison between Rydberg states, harmonic-oscillator Fock states, and the eigenstates of our system is shown in Fig.~\hyperref[fig2]{\ref{fig2}(b)}. We remark that the corresponding Rydberg state for an EDM magnitude of $4$~$e\mu$m would already have a principle quantum number $n\sim 300$, which is quite unstable and unfeasible in experiments. In contrast, the designed trap potential stabilizes the states with a giant EDM and thus makes our platform appealing for quantum applications in the MHz frequency range.

 \begin{figure*}[tb]
    \centering
    \includegraphics{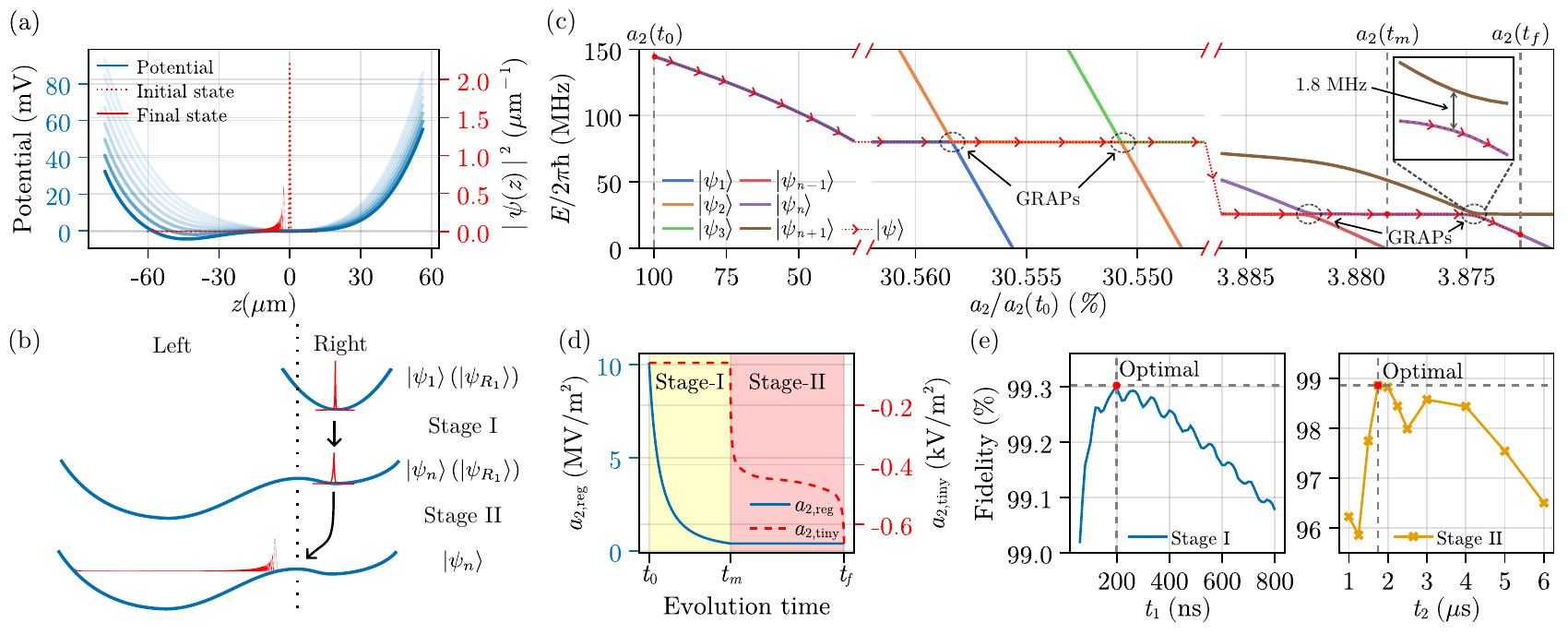}
    \caption{\label{fig3}(a) Potential deformation from the light blue into the deep blue curve. The initial state depicted by the dotted red line is the ground state of the initial potential, whereas the final state depicted by the solid red line is the $3566$th eigenstate of the final potential. (b) Illustration of the evolution throughout the different stages, where $\ket{\psi_{R_1}}$ is the ground state of the right well and $\ket{\psi_j}$ is the $j$th eigenstate of the full potential. (c) Eigenenergies for the eigenstates of the full potential as a function of $a_2$, presenting the ideal trajectory of $\ket{\psi}$ in the $a_2$ parameter space, for example, with $n=3566$.  (e) Fidelity between the ideal eigenstate and the prepared state after stage I and stage II as a function of the evolution time of the two stages. The right panel of (e) is obtained with an optimal stage-I evolution time $t_1=199.4$~ns.}
\end{figure*}

{\it System initialization.---} We proceed to demonstrate that the two-level system formed by the neighboring motional eigenstates with a giant EDM can be efficiently initialized by a potential deformation, an extension of an idea used in a protocol for the preparation of high Fock states of a harmonically trapped ion~\cite{simonTrappedionFockstatePreparation2020}. Initially, the electron is trapped in a standard Paul trap with an axial harmonic potential $a_2 z^2$, where $a_2=m_e\omega_z^{2}/2e$ corresponds to a trap frequency of $\omega_z=(2\pi) 300$~MHz. It is then brought close to the motional ground state using well-established cooling techniques (see SM Sec.~\ref{supp-subsec:cooling}~\cite{supp}). Subsequently, applying additional voltages to the blue DC electrodes, the third- and fourth-order potential shown in Fig.~\hyperref[fig1]{\ref{fig1}(b)} is added. The electron is approximately in the ground state of the new potential, since the higher-order contributions are negligible compared to the initial harmonic potential. The parameter $a_2$ is then gradually decreased to deform the potential, which is represented by the change from a light blue to a deep blue curve in Fig.~\hyperref[fig3]{\ref{fig3}(a)}. At the same time, the state evolves into an eigenstate of the full potential, a component of our two-level system.

The evolution of the state can be divided into two distinct stages, as shown in Fig.~\hyperref[fig3]{\ref{fig3}(b)}. Stage I (from $t_0$ to $t_m$) 
transforms the initial harmonic-potential ground state $\ket{\psi}$ into the $n$th eigenstate $\ket{\psi_n}$ of the full potential. During this stage, the evolution is adiabatic only when $a_2$ is away from GRAPs, where a GRAP represents a specific energy-level anticrossing point in $a_2$ parameter space~\cite{GRAPs}, as seen in Fig.~\hyperref[fig3]{\ref{fig3}(c)}. 
However, at these GRAPs, the state $\ket{\psi}$ non-adiabatically crosses energy levels, resulting in an increment of the quantum number~\cite{zenerNonadiabaticCrossingEnergy1932,wittigLandauZenerFormula2005}.
When the anticrossing gap at the GRAPs is small, stage I can be approximated by an adiabatic decrease of the confinement to the right well and a fast quasiadiabatic method~\cite{martinez-garaotFastQuasiadiabaticDynamics2015} can be utilized uniformly for a speed-up of the process. Thus, one can derive a trajectory of $a_2$ during stage I that obeys
\begin{equation}\label{eq:da2dt}
    \frac{da_2}{dt}=-\frac{\epsilon}{\hbar}\min_{i\neq 1}\abs{\frac{\qty[E_{R_1}\qty(a_2)-E_{R_i}\qty(a_2)]^2}{\mel{\psi_{R_1}\qty(a_2)}{\pdv{H}{a_2}}{\psi_{R_i}\qty(a_2)}}},
\end{equation}
where $\epsilon \ll 1$ is a constant and $E_{R_i}(a_2)$ and $\ket{\psi_{R_i}(a_2)}$ are the $i$th instantaneous eigenvalue and eigenstate, respectively, of the right well for a specific value of $a_2$. Under the assumption that the right well is nearly harmonic, this trajectory reads (see SM Sec.~\ref{supp-subsec:a2 traj}~\cite{supp})
\begin{equation}
a_2(t)=\qty[4\sqrt{e/m_e} \epsilon t+\sqrt{1/a_2(t_0)}]^{-2}.
\end{equation}
Stage II (from $t_m$ to $t_f$) is an adiabatic process, during which the trajectory of $a_2$ can also be calculated according to the fast quasiadiabatic method~\cite{martinez-garaotFastQuasiadiabaticDynamics2015}, yielding a trajectory for the full potential that obeys an equation similar to Eq.~\eqref{eq:da2dt}. An example trajectory is shown in Fig.~\hyperref[fig3]{\ref{fig3}(d)}. The three instances $a_2(t_0)$, $a_2(t_m)$, and $a_2(t_f)$ are also indicated by vertical dashed lines in Fig.~\hyperref[fig3]{\ref{fig3}(c)}. The resulting anticrossing gap at the GRAP between $a_2(t_m)$ and $a_2(t_f)$ is $(2\pi)1.8$~MHz. At $t_f$, the final two-level system is composed of $\ket{\psi_{n}}$ and $\ket{\psi_{n{-}1}}$, with $n=3566$, in this example, possessing a transition frequency of $(2\pi)59.9$~MHz and an EDM magnitude of $7.16$~$e\mu$m.

To realize a precise control of the potential deformation during the two stages, we decompose $a_2 z^2$ into two parts, namely $a_{2,\mathrm{reg}} z^2$ and $a_{2,\mathrm{tiny}} z^2$, generated by the blue and yellow DC electrodes, respectively [cf. Fig.\hyperref[fig1]{\ref{fig1}(a)}]. It is worth noting that, even if $a_{2,\mathrm{tiny}}$ is orders of magnitude smaller than $a_{2,\mathrm{reg}}$, the voltages supplied to the yellow DC electrodes are of the same magnitude as those supplied to the blue DC electrodes, due to the small size and the specific placement of the yellow ones (see SM Sec.~\ref{supp-subsec:DC voltages}~\cite{supp}). In Fig.~\hyperref[fig3]{\ref{fig3}(e)}, we show the fidelity $F$ between the ideal eigenstate $\ket{\psi_{n}}$ and the actually prepared state $\ket{\psi_\mathrm{3D}}$ after stage I (stage II) as a function of the stage-I (stage-II) evolution time $t_1=t_m-t_0$ ($t_2=t_f-t_m$), which is defined as
\begin{equation}\label{eq:projF}
    F=\mel{\psi_\mathrm{3D}}{\qty(I_x\otimes I_y \otimes \dyad{\psi_{n}}) }{\psi_\mathrm{3D}},
\end{equation}
with the identity operators $I_x$ and $I_y$ of the motion in the $x$ and $y$ dimensions, respectively. The left panel of Fig.~\hyperref[fig3]{\ref{fig3}(e)} demonstrates an optimal evolution time of stage I, which balances the competing effects that necessitate fast potential deformation near GRAPs and slow deformation elsewhere. The right panel of Fig.~\hyperref[fig3]{\ref{fig3}(e)} is likewise obtained with this optimal $t_1=199.4$~ns, showing that the initialization fidelity can reach $98.8\%$ for $t_2=1.75$~$\mu$s (see SM Secs.~\ref{supp-subsec:DAC effect init}, \ref{supp-subsec:numerical init}~\cite{supp}).

{\it State readout.---} For the state readout of the two-level system that is composed of the eigenstates $\ket{\psi_n}$ and $\ket{\psi_{n^\prime}}$ of $H_{z}$, we first transfer the information of the motional states onto the spin degree of freedom of the trapped electron using a magnetic-field gradient oscillating with the frequency $\omega_B$ that is resonant with the transition of the two-level system. This is described by the Hamiltonian 
\begin{equation}\label{eq:H_B}
H_B=H_z-\mu_\mathrm{B} b_y  s_{y} z \cos(\omega_B t+\phi_B),
\end{equation}
with the Bohr magneton $\mu_\mathrm{B}$, the $y$ Pauli matrix $s_{y}$ of the spin degree of freedom, and $b_y=\pdv*{B_{y}}{z}$. On the other hand, we denote the $x$ and $y$ Pauli operators of the motional two-level system by $\sigma_{x}$ and $\sigma_{y}$, respectively.
In the interaction picture with respect to $H_{z}$, one can perform a rotating-wave approximation and thereby obtain (see SM Sec.~\ref{supp-subsec:Hamiltonians}~\cite{supp})
\begin{equation}\label{eq:H_B_I}
    H_{B}^{\mathrm{(int)}}=-\frac{1}{2}\hbar g s_{y} \otimes \sigma_{\phi},
\end{equation}
where $\sigma_{\phi}=\cos(\phi)\sigma_{x}+\sin(\phi)\sigma_{y}$, $\phi=\phi_B-\phi_z$, with $\phi_z=\arg(z_{n n^\prime })$, $z_{n n^\prime}=\mel{\psi_{n}}{z}{\psi_{n^\prime}}$, and $g=\mu_\mathrm{B} b_y\abs{z_{n n^\prime}}/\hbar$. 
Hamiltonian~\eqref{eq:H_B_I} induces a rotation of the spin state that is conditional on the motional state. If the spin is initialized into the state $\ket{\uparrow}$ (the $+1$ eigenstate of $s_{z}$), then after $t=\pi/2g$, the spin state will have been rotated into $\ket{+}$ or $\ket{-}$ (namely the $\pm1$ eigenstates of $s_{x}$), depending on the motional-state projection on the eigenstates of $\sigma_{\phi}$, i.e., $\ket{\downarrow_{\phi}}$ or $\ket{\uparrow_{\phi}}$. Therefore, the information on the motional states is transformed into the populations of the spin eigenstates of $s_{x}$, which can then be readout through a spin measurement~\cite{pengSpinReadoutTrapped2017}.

\begin{figure}[t]
    \centering
    \includegraphics{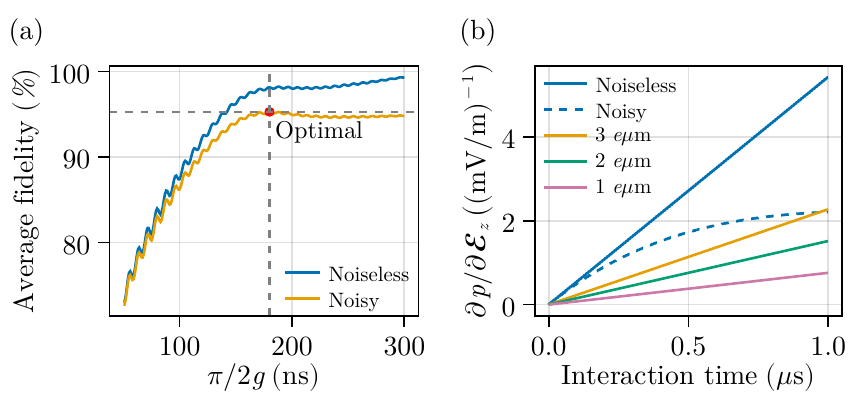}
    \caption{
    (a) Average fidelity as a function of the transfer time $\pi/2g$,
     which is obtained through simulation in the Hilbert space composed by the spin space and the motional subspace formed by the eigenstates $\ket{\psi_{3561}}$ through $\ket{\psi_{3575}}$ of the potential with $a_2(t_f)$ from Fig.~\hyperref[fig3]{\ref{fig3}(c)} (see SM Sec.~\ref{supp-subsec:numerical subspace}~\cite{supp}). (b) Susceptibility $\pdv*{p}{\mathcal{E}_z}|_{\mathcal{E}_z=0}$ as a function of the interaction time for an EDM magnitude of $7.16$~$e\mu$m without (solid blue) and with noise (dashed blue), which is compared with the noiseless results for the EDM magnitudes of $1$ (pink), $2$ (green), and $3$~$e\mu$m (yellow). In both (a) and (b), we assume the spectral density of the electric-field noise to follow the power law $S_E(\omega)\propto 1/\omega^{1.3}$ with $S_E[(2\pi)1$~MHz$]\approx10^{-12}$~(V/m)$^2$/Hz for a $4$-K environment and a particle-surface distance of $30$~$\mu$m~\cite{brownnuttIontrapMeasurementsElectricfield2015,sedlacekEvidenceMultipleMechanisms2018}.}
    \label{fig4}
\end{figure}

We proceed to numerically simulate this transfer process and calculate the average fidelity $F_\mathrm{avg}=\qty(F_{+}+F_{-})/2$, where $F_\pm$ represents the fidelity between the final spin state and the ideal state $\ket{\pm}$ for the initial motional states $\ket{\downarrow_{\phi}}$ and $\ket{\uparrow_{\phi}}$, respectively. The influence of electric-field noise is taken into account by employing a master equation (see SM Sec.~\ref{supp-subsec:surface noise}~\cite{supp}), where the noise-induced transition rate between the eigenstates is specified by the spectral density of the noise. Figure~\hyperref[fig4]{\ref{fig4}(a)} shows how the average fidelity changes with the spin-motion coupling strength $g$. If the coupling is too strong to ensure the validity of the rotating-wave approximation, the averaged fidelity would be very low. On the other hand, if the coupling is too week, the averaged fidelity will also decrease due to the action of the noise over the long transfer time $\pi/2g$.
As shown in Fig.~\hyperref[fig4]{\ref{fig4}(a)}, the average fidelity can reach $95.2\%$ for an optimal transfer time of $180$~ns, which requires a magnetic-field gradient $b_y=0.14$~G$\cdot\mu$m$^{-1}$ that is experimentally feasible~\cite{ospelkausMicrowaveQuantumLogic2011,srinivasTrappedIonSpinMotionCoupling2019}.

{\it Coherent control and quantum sensing.---} Owing to the nonlinear eigenenergies, the artificial atomic dipoles can be coherently controlled as a two-level system by an oscillating electric field with amplitude $\mathcal{E}_z$, frequency $\omega_E$, and phase $\phi_E$. 
This can be described by the Hamiltonian
\begin{equation}\label{eq:electric_H}
    H_E=H_z-ez\mathcal{E}_z\cos(\omega_E t+\phi_E)
\end{equation}
 and the underlying mechanism is similar to quantum sensing of weak electric fields. To characterize how fast the two-level system formed by $\ket{\psi_n}$ and $\ket{\psi_{n^\prime}}$ can be controlled, we introduce the effective detuning
\begin{equation}\label{eq:detune}
\Delta\omega_{nn^\prime}=\min_{i\in(n,n^\prime),j\notin(n,n^\prime)}\left(\frac{\abs{\mu_{n n^\prime}}}{\abs{\mu_{ij}}}\Delta\omega_{ij,nn^\prime}\right),
\end{equation}
where $\Delta\omega_{ij,nn^\prime}=\abs{\abs{\omega_{ij}}-\abs{\omega_{nn^\prime}}}$ with $\omega_{ij}=(E_i-E_j)/\hbar$. The coherent-control Rabi frequency $\Omega_R=\abs{\mu_{nn^\prime}} \mathcal{E}_z/\hbar$ is required to be much smaller than the effective detuning to ensure that leakage to other energy levels is negligible (see SM Sec.~\ref{supp-subsec:Hamiltonians}~\cite{supp}). For $a_2(t_f)$ from Fig.~\hyperref[fig3]{\ref{fig3}(c)}, one finds an effective detuning of $(2\pi)5.5$~MHz and a transition frequency of $(2\pi)59.9$~MHz for the two-level system composed of $\ket{\psi_{3566}}$ and $\ket{\psi_{3565}}$. A larger effective detuning could be achieved by choosing $a_2(t_f)$ near a GRAP, which would lead to a smaller corresponding transition frequency (see SM Sec.~\ref{supp-subsec: a2tm and a2tf}~\cite{supp}).

Lastly, in the same framework, we consider the quantum sensing of a weak electric field along the axial direction that is oscillating on resonance with the transition of the two-level system, i.e., with the frequency $\omega_E=\omega_{nn^\prime}$. The system is initialized into $\ket{\psi_n}$, and after an interaction time $t$, the probability to find the system in the $+1$ eigenstate of $\sigma_{\phi}$, with $\phi=\phi_E-\phi_z -\pi/2$, reads $p=[1+\sin(\Omega_R t)]/2$ (see SM Sec.~\ref{supp-subsec:Hamiltonians}~\cite{supp}). 
Taking into account the influence of the electric-field noise, the susceptibility of this probability to the electric field strength can be written as $\frac{\partial p}{ \partial \mathcal{E}_z }|_{\mathcal{E}_z=0}=\abs{\mu_{nn^\prime}}\exp(-\Gamma t)t/2\hbar$,
where $\Gamma$ is the effective decay rate (see Refs.~\cite{suterColloquiumProtectingQuantum2016a,degenQuantumSensing2017} and SM Sec.~\ref{supp-subsec:numerical subspace}~\cite{supp}).
Figure~\hyperref[fig4]{\ref{fig4}(b)} demonstrates that, even in a noisy environment, the giant-EDM sensor presents an improved susceptibility in a certain range of the interaction time, indicating an enhanced metrological performance~\cite{chuDynamicFrameworkCriticalityEnhanced2021}.

{\it Conclusion.---} We have presented a novel method to create two-level systems with a giant EDM, that are formed by motional states of an electron confined in a specially engineered Paul trap. In order to demonstrate the practicality of the proposed method, we have presented the efficient initialization and read out, as well as the coherent manipulation of the system. Our detailed analysis and numerical simulations, taking into account realistic experimental conditions, ensure the feasibility of our approach with the state-of-the-art experimental capabilities.
Furthermore, we have illustrated a simple protocol for electric field sensing, showcasing a very prominent susceptibility to the electric field strength. Our work represents a promising approach to create giant EDMs in artificial quantum systems and opens appealing possibilities for coherent atomic control and quantum technologies. 
The design principles of our approach could potentially be extended to other trapped-electron systems with flexible controllability of the electric potential, such as quantum dots \cite{scarlinoSituTuningElectricDipole2022}, helium- or neon-trapped electrons~\cite{koolstraCouplingSingleElectron2019,zhouSingleElectronsSolid2022}.
Moreover, the extension to multi-electron scenarios would provide an interesting platform for the study of interacting electrons.

{\it Acknowledgements.---}This work is supported by the National Natural Science Foundation of China (Grants No.~12161141011 and No.~12174138), the National Key R$\&$D Program of China (Grant No. 2018YFA0306600), and the Shanghai Key Laboratory of Magnetic Resonance (East China Normal University). Y.-M.C. is also supported by the fellowship of China Postdoctoral Science Foundation (Grant No. 2022M721256). Part of the computation was completed on the HPC Platform of Huazhong University of Science and Technology.

\nocite{suterColloquiumProtectingQuantum2016a,aminiMicrofabricatedChipTraps2008,hongGuidelinesDesigningSurface2016,yuFeasibilityStudyQuantum2022,jacksonClassicalElectrodynamics1999,penroseBestApproximateSolutions1956,pengSpinReadoutTrapped2017,gormanTwomodeCouplingSingleion2014,martinez-garaotFastQuasiadiabaticDynamics2015,comparatDipoleBlockadeCold2010,suzukiFractalDecompositionExponential1990,besardEffectiveExtensibleProgramming2019,degenQuantumSensing2017,brownnuttIontrapMeasurementsElectricfield2015,savardLasernoiseinducedHeatingFaroff1997,kramerQuantumOpticsJlJulia2018,rackauckasDifferentialEquationsJlPerformant2017,whiteJuliaDiffFiniteDifferencesJl2022,brownMaterialsChallengesTrappedion2021,sedlacekEvidenceMultipleMechanisms2018,vandrielFrequencyDependentSpontaneousEmission2005}

\bibliography{GiantDipole}

\begin{thebibliography}{87}
\bibitem [{GRA()}]{GRAPs}%
  \BibitemOpen
  \href@noop {} {\bibinfo  {journal} {At GRAPs, one of the eigenstates forming
  the energy-level anticrossing is initially similar to the ground state of the
  right well. "GR" in GRAP therefore represents "Ground state of the Right
  well" and "AP" stands for "Anticrossing Point"}\ }\BibitemShut {NoStop}%
\end{thebibliography}

 

\section*{Supplementary material}

# Supplemental Material:

# Engineering artificial atomic systems of giant electric dipole moment

Baiyi Yu,[1] Yaoming Chu,[1, *] Ralf Betzholz,[1] Shaoliang Zhang,[1, †] and Jianming Cai[1, 2, ‡]

[1]*School of Physics, International Joint Laboratory on Quantum Sensing and Quantum Metrology,*

*Hubei Key Laboratory of Gravitation and Quantum Physics,*

*Institute for Quantum Science and Engineering, Wuhan National High Magnetic Field Center,*

*Huazhong University of Science and Technology, Wuhan 430074, China*

[2]*Shanghai Key Laboratory of Magnetic Resonance, East China Normal University, Shanghai 200062, China*

(Dated: April 20, 2023)

## CONTENTS

---

* yaomingchu@hust.edu.cn

† shaoliang@hust.edu.cn

‡ jianmingcai@hust.edu.cn



## I. ENGINEERING OF THE ELECTRIC POTENTIAL

In this section, we first give a complete and detailed description on our trap geometry. Then, we discuss the RF trapping in radial directions ($x$ and $y$ directions). After that, we give a detailed derivation of the DC-potential parameters $a_3$ and $a_4$ shown in Eq. (2) of the main text, and a detailed derivation of the effective axial potential described by Eq. (3) of the main text. Finally, we discuss the numerical details on calculating the DC-electrode voltages for different objective potentials, and the effect of digital-to-analog converter (DAC) on the DC-electrode voltages.

### A. Trap geometry

The prototype for our trap is composed of two identical layers of electrodes, which are separated by 60 $\mu$m along the $y$ direction, see Fig. S1 for a schematic representation. The two different phases, with a phase difference of $\pi$, of the radio-frequency (RF) voltages that are applied to the four red electrodes are represented by $\oplus$ and $\ominus$. A top view showing the detailed electrode geometry is shown in Fig. S2. In the case of the blue and yellow DC electrodes, there are always four electrodes with same $z$-coordinate which are supplied with the same voltage, possess the same shape, and are therefore labeled identically. All electrodes are separated from one another by a 2 $\mu$m gap along the $z$ direction. Table S1 presents a comprehensive description of the electrode dimensions. Here, the dimensions of the electrodes 6 through 11 correspond to their rectangular shape prior to the consideration of the electrodes 16 through 24. The actual geometries of electrodes 6 through 11 are obtained by subtracting the shapes of the electrodes 16 through 24 and the intervening gaps from the original rectangles.

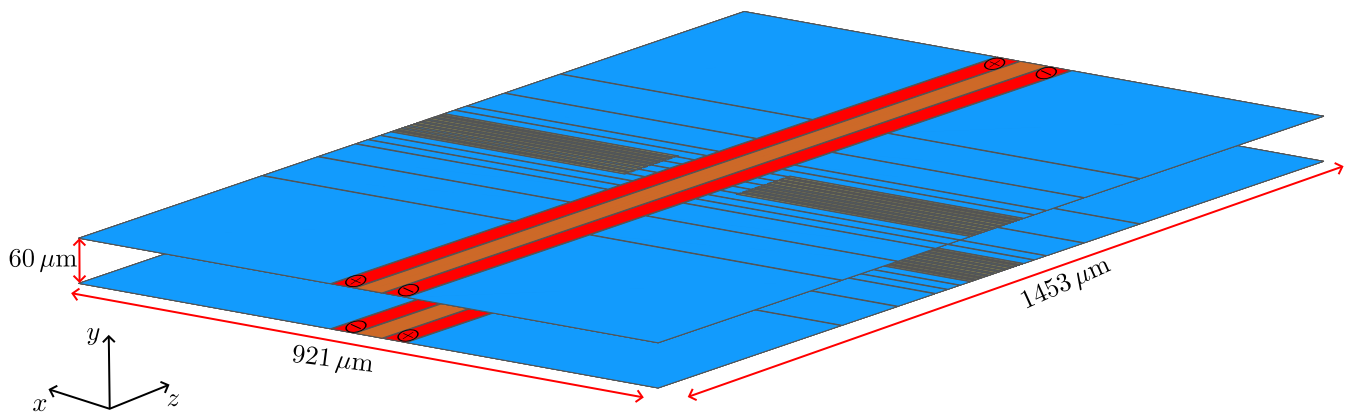

FIG. S1. Schematic of our trap prototype with two electrode layers separated by 60 $\mu$m.

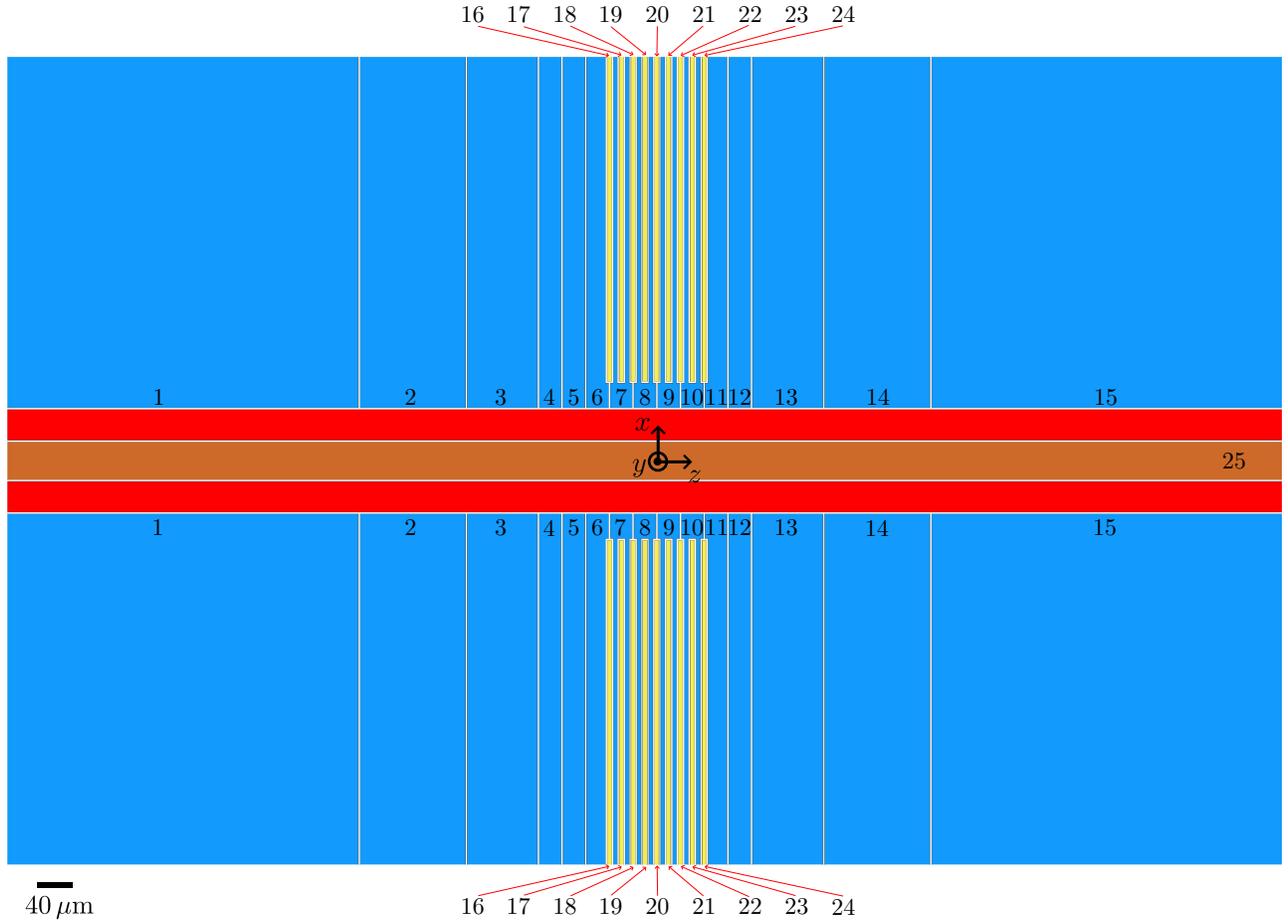

FIG. S2. Complete electrode geometry. Blue, yellow, and brown represent DC electrodes whereas red denotes RF electrodes.

| Number | RF | 1 | 2 | 3 | 4 | 5 | 6 | 7 | 8 | 9 | 10 | 11 | 12 | 13 | 14 | 15 | 16 | 17 | 18 | 19 | 20 | 21 | 22 | 23 | 24 | 25 |
|---|---|---|---|---|---|---|---|---|---|---|---|---|---|---|---|---|---|---|---|---|---|---|---|---|---|---|
| $W$ ($\mu$m) | 35 | 400 | 120 | 80 | 25 | 25 | 25 | 25 | 25 | 25 | 25 | 25 | 25 | 80 | 120 | 400 | 5 | 5 | 5 | 5 | 5 | 5 | 5 | 5 | 5 | 43 |
| $L$ ($\mu$m) | 1453 | 400 | 400 | 400 | 400 | 400 | 400 | 400 | 400 | 400 | 400 | 400 | 400 | 400 | 400 | 400 | 370 | 370 | 370 | 370 | 370 | 370 | 370 | 370 | 370 | 1453 |

TABLE S1. Electrode dimensions. Width and length are denoted by $W$ and $L$, respectively.

### B. Pseudopotential of RF field and trapping stability

In our design, the RF field is generated by supplying the AC electrodes [red in Fig. S1 and Fig. S2] with RF voltages. When the RF frequency $\omega_{\mathrm{RF}}$ is much larger than the secular frequency $\omega_r$, the effect of RF field can be approximately described by the pseudopotential $\Phi_{\mathrm{pp}}$, which can be calculated by

$$\Phi_{\mathrm{pp}}(\mathbf{r}) = \frac{e|\mathbf{E}(\mathbf{r})|^2}{4m\omega_{\mathrm{RF}}^2}, \tag{S1}$$

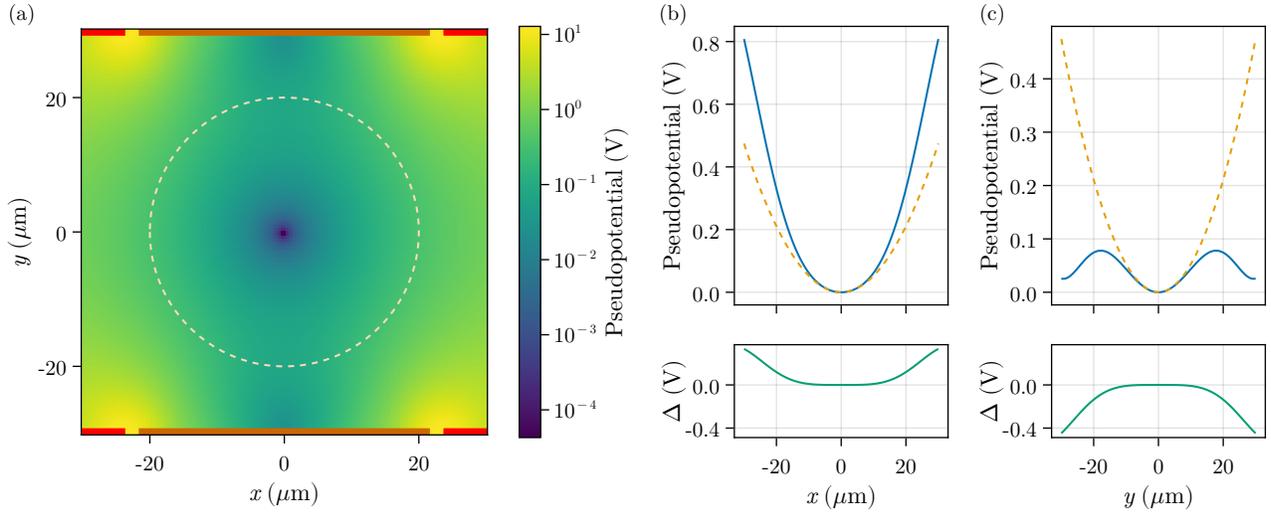

FIG. S3. (a) Pseudopotential for $z = 0$. Brown and red lines represent the central DC electrodes and the RF electrodes, respectively. (b)-(c) Top panels: Solid blue lines represent the pseudopotential along the $x$ and $y$ axes, whereas dashed yellow lines show the ideal harmonic potential with frequency $(2\pi)2.17$ GHz. Bottom panels: Difference $\Delta$ between the pseudopotential and the ideal harmonic potential.

where $\mathbf{E}(\mathbf{r})$ is the RF electric field at position $\mathbf{r}$ [1, 2]. In Fig. S3, we depict this pseudopotential produced by our trap with an RF drive of frequency $\omega_{\mathrm{RF}} = (2\pi)50$ GHz and voltage $U_{\mathrm{RF}} = 64$ V with all DC electrodes being grounded. If we consider a DC potential that generates a $(2\pi)300$ MHz harmonic confinement in the axial direction ($z$ direction), the secular frequency $\omega_r$ in the radial directions is expected to be around $(2\pi)2.15$ GHz.

To assess the stability of the trapping potential, we perform simulations of classical electron trajectories in the $x$-$y$ plane, with the RF drive activated and the DC electrodes grounded, for various initial phases of the RF drive and different initial energies of the electron in the RF potential. The free electrons are assumed to be created through photoionization. Their initial energy is only related to the distance from the position where the ionization occurs to the trap center, since the surplus kinetic energy from the ionization is negligible in comparison with the potential energy [3]. Figure S4 displays a map of the electron storage time derived from the simulation results. The simulation time extends up to 100 $\mu$s, and we consider an electron to be lost once its distance to the origin $x = y = 0$ exceeds 20 $\mu$m [the region enclosed by a dashed circle in Fig. S3(a)]. As shown in Fig. S4, for ionization positions smaller than 11 $\mu$m, the electron trajectories are stable under various phases of the RF drive (region to the right of the dashed line). However, for larger ionization positions, the stability strongly depends on the phase of the RF drive, and the electron trajectories exhibit the greatest stability when the phase of the RF drive is zero.


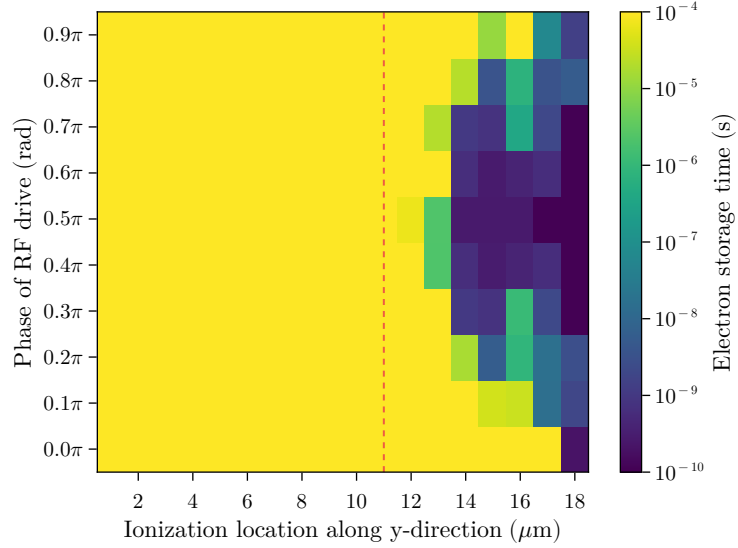

FIG. S4. Dependence of the electron storage time on the phase of the RF drive and the $y$ coordinate of the ionization.

### C. Derivation of DC-potential parameters $a_3$ and $a_4$

Equation (2) of the main text introduces two new parameters, $d$ and $a'_2$, to calculate $a_3$ and $a_4$. Here, $d$ represents the distance between the two points that satisfy

$$\frac{\partial(a_3 z^3 + a_4 z^4)}{\partial z} = 0. \tag{S2}$$

This directly yields $3a_3 z^2 + 4a_4 z^3 = 0$, which has the double root $z = 0$ as well as the real root $z = -3a_3/4a_4$. Since $a_3, a_4 > 0$, we have

$$d = \frac{3a_3}{4a_4}. \tag{S3}$$

On the other hand, $a'_2$ is the second-order coefficient of the Taylor expansion of $a_3 z^3 + a_4 z^4$ around $z = -d$, describing the approximately harmonic confinement of the potential $a_3 z^3 + a_4 z^4$ around $z = -d$. We thereby have

$$\frac{12 a_4 d^2 - 6 a_3 d}{2} = a'_2, \tag{S4}$$

and by combining Eq. (S3) with Eq. (S4) we obtain $a_3$ and $a_4$ shown in Eq. (2) of the main text. If we denote $\omega'_h$ as this approximately harmonic frequency, we have $a'_2 = m_e {\omega'_h}^2 / 2e$.

## D. Derivation of effective axial potential

The desired 1D potential takes the form

$$\Phi_{1D} = a_2 z^2 + a_3 z^3 + a_4 z^4, \tag{S5}$$

where $a_2 z^2$ is related to the initialization process. However, in a 3D space free of charge, static electric potentials with $z^3$ and $z^4$ terms cannot be generated without introducing coupling terms between axial and radial motion, such as $xz^2$ and $y^2 z^2$, based on the Laplace equation of static electric potential [4]. Under radial symmetry, the 3D potential with terms of $\Phi_{1D}$ can be expressed as

$$\begin{aligned}\Phi_{3D} =\, & a_2(z^2 - \frac{1}{2}x^2 - \frac{1}{2}y^2) + a_3(z^3 - \frac{3}{2}zx^2 - \frac{3}{2}zy^2) \\ & + a_4(z^4 - 3z^2 x^2 - 3z^2 y^2 + \frac{3}{8}x^4 + \frac{3}{8}y^4 + \frac{3}{4}x^2 y^2),\end{aligned} \tag{S6}$$

which is composed of spherical harmonics. When the confinement in the radial directions is much larger than the coupling coefficients, the coupling terms in Eq. (S6) can be treated as axial correction terms, resulting in an effective axial potential.

To illustrate this clearly, we consider the potential in the Hamiltonian to be composed of the DC potential (S6) and an RF pseudopotential which acts as harmonic potentials in the $x$ and $y$ directions. In the quantum regime, we can rewrite the coupling terms in Eq. (S6) with $x = \sqrt{\hbar/2m\omega_r}(a_x^\dagger + a_x)$ and $y = \sqrt{\hbar/2m\omega_r}(a_y^\dagger + a_y)$, where $a_x$ and $a_y$ are the annihilation operators in the $x$ and $y$ directions, respectively. As an example, $zx^2$ can be expressed as

$$zx^2 = \frac{\hbar}{2m\omega_r} z\big(a_x^{\dagger 2} + a_x^2 + 2a_x^\dagger a_x + 1\big). \tag{S7}$$

In the interaction picture, $a_x^{\dagger 2}$ and $a_x^2$ are rapidly-oscillating terms and the coupling coefficient $\Omega$ related to the eigenstates $|\psi_n\rangle$ and $|\psi_{n'}\rangle$ of the uncorrected Hamiltonian in the $z$ direction can be written as

$$\Omega = \left| \frac{3}{2} e a_3 \frac{\hbar}{2m\omega_r} \langle \psi_n | z | \psi_{n'} \rangle \right|. \tag{S8}$$

Typically, we assume $\langle \psi_n | z | \psi_{n'} \rangle = 5$ $\mu$m and $\omega_r = 2$ GHz and $a_3$ is determined by Eq. (2) of the main text with $d = 40$ $\mu$m and $\omega_h' = (2\pi)300$ MHz. In this case, we have $\Omega \approx 8.8$ MHz, which is much smaller than $\omega_r$. Applying rotating-wave approximation, $zx^2$ can be written as

$$zx^2 \approx \frac{\hbar}{2m\omega_r} z\big(2a_x^\dagger a_x + 1\big). \tag{S9}$$

Considering that the electron is in the ground state of the motion in the $x$ direction, we can substitute $a_x^\dagger a_x$ with zero and get

$$zx^2 \approx \frac{\hbar}{2m\omega_r} z, \tag{S10}$$

which is merely a correction to the axial potential. Similarly, we can obtain the coupling coefficient of the $z^2x^2$ term, which is around $0.4$ MHz for the above parameters, and similarly those of the terms containing $y$. Then, by applying the rotating-wave approximation we arrive at their effective form. Furthermore, if we revisit Eq. (S10), we find

$$zx^2 \approx z \left\langle x^2 \right\rangle, \tag{S11}$$

which is also suitable for other coupling terms such as $z^2x^2$ or $z^2y^2$ as long as the rotating-wave approximation is valid and the state of the electron in the radial directions is in one of the eigenstates. Small correction to radial potential are caused by $x^4$ and $y^4$ terms. However, they are a second-order corrections to the axial potential and thereby negligible. Furthermore, the $x^2y^2$ term are also negligible as long as the electron has the same phonon number in the $x$ and $y$ directions. Therefore, we can simply add average correction terms to the axial potential to obtain the effective potential in the axial direction, which has the form

$$\Phi_c(z) = \iint \Phi_{3D}(x,y,z)|\psi(x)|^2|\psi(y)|^2 \, dx \, dy, \tag{S12}$$

where $\psi(x)$ and $\psi(y)$ represent the electron wave functions in the $x$ and $y$ directions, respectively. Equation (S12) not only accounts for the interaction terms introduced theoretically, but also for the imperfections of DC potential arising from realistic experimental conditions. Moreover, we can integrate the imperfections of the static RF pseudopotential, $\Phi_{\text{pp}} - \Phi_{\text{pp}}^{\text{ideal}}$ into Eq. (S12), if the imperfections of the static RF pseudopotential also satisfy the assumptions above. Since $\iint \Phi_{\text{pp}}^{\text{ideal}}|\psi(x)|^2|\psi(y)|^2 \, dx \, dy$ is a constant, we can replace $\Phi_{\text{pp}} - \Phi_{\text{pp}}^{\text{ideal}}$ with $\Phi_{\text{pp}}$, and obtain Eq. (3) of the main text.

### E. Numerical details on calculating DC-electrode voltages

It is quite challenging for the present DC electrode design to produce a DC potential that is precisely described by Eq. (S6), for $d = 40$ $\mu$m and $\omega'_h = (2\pi)300$ MHz. Although a reduction of the separation between the two layers could be advantageous, it would also increase the surface noise and decrease the trapping efficiency. Fortunately, the system initialization and the giant EDM do not necessitate a potential that exactly possesses the third-order and fourth-order terms described in Eq. (S6). It is only required that the effective potential in the axial direction is similar to the one described by Eq. (S5).

In our trap design, we directly compute the electrode voltages by minimizing an objective function that represents the difference between the generated potential and the objective potential. Let $\mathbf{V}_i$ represent the potential generated by the $i$th electrode supplied with a 1-V voltage, while all other electrodes are grounded. We represent the matrix composed of $\mathbf{V}_i$ as its columns by $\mathbf{V}_M$ and denote by $\mathbf{c}$ a vector whose $i$th component, $c_i$, represents the voltage supplied to the $i$th electrode. The generated potential can then be expressed as $\mathbf{V}_M\mathbf{c} = \sum_i c_i \mathbf{V}_i$ [4], and we can choose the objective function $f_{\text{obj}}$ to have the form

$$f_{\text{obj}}(\mathbf{c}) = \|\mathbf{V}_M\mathbf{c} - \mathbf{V}_{\text{obj}}\|_2, \tag{S13}$$

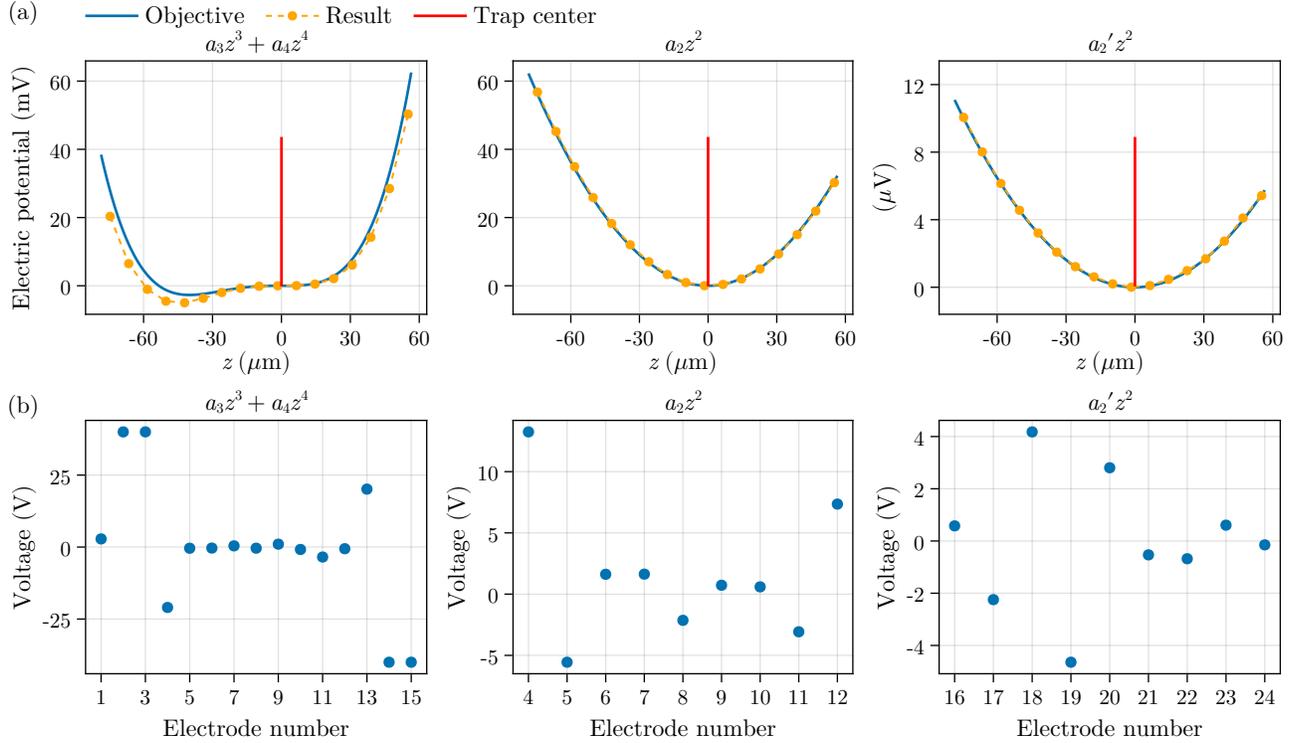

FIG. S5. (a) Simulated potential at $x = y = 0$: $a_3 = 1.68 \times 10^{11}$ V/m$^3$ and $a_4 = 3.16 \times 10^{15}$ V/m$^4$ determined by Eq. (2) of the main text with $d = 40$ $\mu$m and $\omega' = (2\pi)300$ MHz; $a_2 = 1.01 \times 10^7$ V/m$^2$ corresponding to $\omega_z = (2\pi)300$ MHz and $a_2' = 1800$ V/m$^2$. (b) Voltages supplied to the electrodes to generate the potential in (a). Electrodes with numbers not listed are grounded.

where $\|\mathbf{x}\|_2 = \sqrt{\sum_i x_i^2}$ is the $L^2$ norm and vector $\mathbf{V}_{\text{obj}}$ is the objective potential. Since adding a constant to the potential does not affect the motion of the electron, we can include a virtual offset $c_c \mathbf{V}_c$, with $\mathbf{V}_c$ being the vector whose components are all equal to 1, into Eq. (S13) by appending $c_c$ as an additional component to the vector $\mathbf{c}$ and $\mathbf{V}_c$ as an additional column to the matrix $\mathbf{V}_M$. By minimizing $f_{\text{obj}}(\mathbf{c})$ under different objective potentials and adhering to constraints on the voltages supplied to the electrodes, we can obtain multiple sets of $\mathbf{c}$. If an unconstrained minimization is preferred, the Moore-Penrose pseudoinverse [5] can be utilized to obtain the $\mathbf{c}$ with $\mathbf{c} = \text{pinv}(\mathbf{V}_M) \cdot \mathbf{V}_{\text{obj}}$, where pinv is a function returning Moore-Penrose pseudoinverse. Furthermore, considering the electrode symmetry, we can speed up the calculation by replacing the 3D objective potential with a potential that solely accounts for the potentials in the $z$ and $x$ directions.

The results are shown in Fig. S5, where the potentials at $z = 0$ are adjusted to 0 by a constant, allowing for a clearer comparison. The left panels of Fig. S5 are obtained with the constraints $-40$ V $< c_i <$ 40 V, the central panels are obtained with the constraints $-20$ V $< c_i <$ 20 V, and the right panels are obtained using the pseudoinverse method. It is worth noting that, after minimizing $f_{\text{obj}}(\mathbf{c})$ for objective potential of the form $a_3 z^3 + a_4 z^4$ at $x = y = 0$, we add the electrode voltage vector

c with c′, which generates appropriate first and second order potential to cancel the first-order and second-order terms of the generated potential at $z = 0$, to make the generated potential as close to the objective potential at trap center as possible.

### F. Effect of digital-to-analog converter on DC-electrode voltages

In real experiment, we need to consider the effect of the digital-to-analog converter (DAC). Denoting the voltage step of the $i$th electrode as $c_i^{\text{step}}$, for a 16-bit DAC, $c_i^{\text{step}} \times 2^{16}$ should be larger than the required range of $c_i$ during the whole process. The voltage steps we finally choose are shown in Tab. S2 (not covering the GND electrode, whose voltage is set to 0 during the whole procedure).

| Number | 1 | 2 | 3 | 4 | 5 | 6 | 7 | 8 | 9 | 10 | 11 | 12 | 13 | 14 | 15 | 16 | 17 | 18 | 19 | 20 | 21 | 22 | 23 | 24 |
|---|---|---|---|---|---|---|---|---|---|---|---|---|---|---|---|---|---|---|---|---|---|---|---|---|
| Voltage step ($\mu$V) | 1200 | 1200 | 1200 | 1200 | 300 | 60 | 180 | 180 | 60 | 30 | 300 | 300 | 1200 | 1200 | 1200 | 15 | 60 | 120 | 120 | 60 | 15 | 15 | 15 | 3 |

TABLE S2. Voltage steps of DC electrodes.

## II. DETAILS ON SYSTEM INITIALIZATION

In this section, we first discuss details on the initial-state cooling. Then, we discuss the requirements on the values of $a_2$ at the end points of stage I and stage II, $a_2(t_m)$ and $a_2(t_f)$, respectively, which are of vital importance for the system initialization. Next, we give calculation details on the trajectory of $a_2$. After that, we discuss the effect of DAC, before we finally give numerical details on the state evolution during the system initialization.

### A. Details on initial-state cooling

The system initialization requires an initial cooling of the system to the ground state of its motion. Cooling the radial motion is possible by coupling a cryogenic tank to the system through an image current induced by cutting one of the GND electrodes into two pieces [3]. However, directly cooling the axial motion is more challenging due to its lower frequency. We can follow Ref. [6] to couple the axial motion to the radial one, making the axial temperature $T_z = \omega_z T_r / \omega_r$, which has been realized in ion traps [7].

For a harmonic oscillator with frequency $\omega$ and annihilation operator $a$, the density matrix of a thermal state at temperature $T$ has the form $\rho = [\bar{n}/(\bar{n}+1)]^{a^\dagger a}/(\bar{n}+1)$, with the mean thermal phonon number $\bar{n} = [\exp(\hbar\omega/k_B T) - 1]^{-1}$. The population of the ground state is thereby given by $p_0 = (\bar{n}+1)^{-1}$. Table S3 shows $\bar{n}$ and $p_0$ for different temperatures and harmonic-

| $T$ (K) | 4 | 0.4 | 0.04 | 4 | 0.4 | 0.04 |
|---|---|---|---|---|---|---|
| $\omega/2\pi$ (GHz) | 2 | 2 | 2 | 5 | 5 | 5 |
| $\bar{n}$ | 41.2 | 3.7 | 0.1 | 16.2 | 1.2 | 0.002 |
| $p_0$ | 0.02 | 0.21 | 0.909 | 0.06 | 0.45 | 0.998 |

TABLE S3. Mean thermal phonon number $\bar{n}$ and ground-state population $p_0$ of a harmonic oscillator for different temperatures $T$ and harmonic frequencies $\omega$.

oscillator frequencies. In our case, for the radial secular frequency $\omega_r = (2\pi)2$ GHz, we have a ground-state population $p_{0z} = p_{0r} = 90.9\%$ for a tank temperature $T_t = 40$ mK. The cooling can be either enhanced by decreasing the tank temperature or enlarging the radial secular frequency as shown in Tab. S3.

### B. Requirements on $a_2(t_m)$ and $a_2(t_f)$

Since the state is expected to non-adiabatically cross the energy levels at GRAPs during stage I, $a_2(t_m)$ is naturally required to be away from GRAPs to maximize the fidelity after the stage-I evolution, which can be easily achieved when the anticrossing gap of the GRAP near $a_2(t_m)$ is small. A small anticrossing gap of the GRAP near $a_2(t_m)$ also ensures the attainability of non-adiabatic crossings in stage I. However, a small anticrossing gap of the GRAP near $a_2(t_m)$ results in a long evolution time in stage II, which is not preferable. In practice, we need to balance these competing demands to choose a proper $a_2(t_m)$. In the range of $a_2$ we considered, the smaller $a_2(t_m)$ is, the larger anticrossing gap of the GRAP near $a_2(t_m)$ will be, due to the lower barrier potential between the left and right well. In Fig. 3(c) of the main text, $a_2(t_m) = a_{2,\text{reg}}(t_m) + a_{2,\text{tiny}}(t_m)$, where $ea_{2,\text{reg}}(t_m)\tilde{L}^2/\tilde{E} = 5.142$ with $\tilde{L} = 1$ μm and $\tilde{E} = \hbar^2/(m_e\tilde{L}^2)$, and $a_{2,\text{tiny}}(t_m) = -56.1$ V/m². The anticrossing gap of the GRAP near such $a_2(t_m)$ is $(2\pi)1.8$ MHz.

As for $a_2(t_f)$, one can choose it either away from GRAPs or near a GRAP. For $a_2(t_f)$ near a GRAP, the two-level system is formed by the eigenstates composing the anticrossing at this GRAP, whereas, for $a_2(t_f)$ away from GRAPs, the system is formed by the eigenstates mainly distributed in the left well. Figure S6 and Table S4 illustrates the differences in the magnitude of the EDM, the resonance frequency, and the effective detuning between the two choices of $a_2(t_f)$. The magnitudes of EDMs for both choices are of the same order. However, the resonance frequency and effective detuning with $a_2(t_f)$ near a GRAP are distinct from those with $a_2(t_f)$ away from GRAPs. Specifically, the resonance frequency with $a_2(t_f)$ near a GRAP is much smaller than that with $a_2(t_f)$ away from GRAPs. This is because the resonance frequency with $a_2(t_f)$ near a GRAP equals the anticrossing gap of the GRAP, which is considerably smaller than other normal energy-level gaps. In the same way, the effective

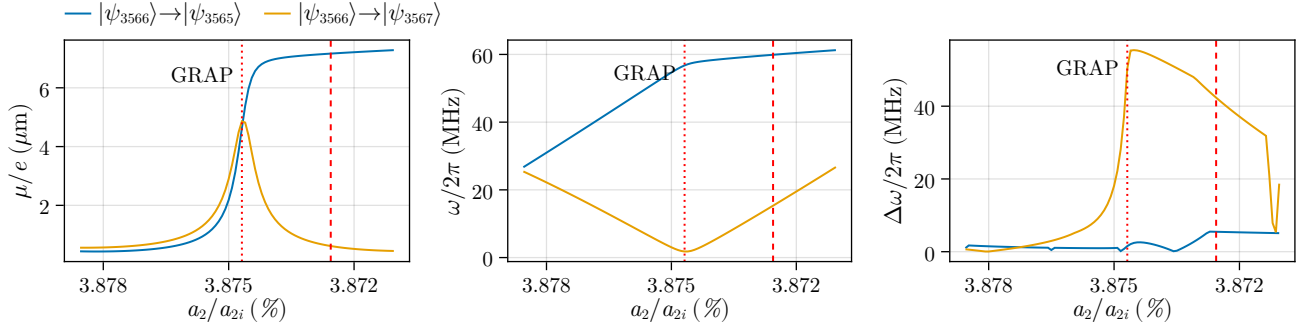

FIG. S6. Magnitude of the EDM $\mu$, resonant frequency $\omega$, and effective detuning $\Delta\omega$ as functions to $a_2$. The panels clearly show the main differences between the two choices of $a_2(t_f)$, which are indicated by the dotted (near GRAP) and dashed (away from GRAP) red lines. The corresponding anticrossing levels are $|\psi_{3566}\rangle$ and $|\psi_{3567}\rangle$.

| Type | $ea_{2,\text{reg}}(t_f)\tilde{L}^2/\tilde{E}$ | $a_{2,\text{tiny}}(t_f)$ (V/m$^2$) | Eigenstates | $\mu/e$ ($\mu$m) | $\omega$ (MHz) | $\Delta\omega$ (MHz) |
|---|---|---|---|---|---|---|
| Away from GRAP | 5.142 | -665.8 | $|\psi_{3565}\rangle \to |\psi_{3566}\rangle$ | 7.16 | 59.9 | 5.5 |
| Close to GRAP | 5.142 | -451.1 | $|\psi_{3566}\rangle \to |\psi_{3567}\rangle$ | 4.86 | 1.8 | 50 |

TABLE S4. Magnitude of the EDM $\mu$, resonance frequency $\omega$, and effective detuning $\Delta\omega$ for different choices of $a_2(t_f)$. We define the length unit $\tilde{L} = 1$ $\mu$m and the energy unit $\tilde{E} = \hbar^2/(m_e\tilde{L}^2)$ to make the values of $ea_{2,\text{reg}}$ precisely match those we use in simulations.

detuning with $a_2(t_f)$ near a GRAP is much larger than that with $a_2(t_f)$ away from GRAPs. Furthermore, the stage-II evolution time with $a_2(t_f)$ near a GRAP is shorter than that with $a_2(t_f)$ away from GRAPs due to the shorter path in $a_2$ parameter space.

### C. Calculation details on $a_2$ trajectory

The trajectory of $a_2$ shown in Fig. 3(d) of the main text is calculated following the fast quasiadiabatic (FAQUAD) approach [8]. With FAQUAD, we have

$$\frac{da_2}{dt} = -\frac{\epsilon}{\hbar}\min_{i \neq n}\left|\frac{[E_n(a_2) - E_i(a_2)]^2}{\langle\psi_n(a_2)|\frac{\partial H}{\partial a_2}|\psi_i(a_2)\rangle}\right|, \tag{S14}$$

where $\epsilon \ll 1$ is a constant and $E_i$ as well as $|\psi_i\rangle$ have different meanings for different stages. During stage I, from $t = 0$ to $t = t_m$, $E_i$ and $|\psi_i\rangle$ can be denoted by $E_{R_i}$ and $|\psi_{R_i}\rangle$, with $i$ starting from 1, respectively representing the $i$th instantaneous eigenvalue and eigenstate of the right well. Furthermore, here, we have $n = 1$. On the other hand, during stage II, from $t = t_m$ to $t = t_f$, $E_i$ and $|\psi_i\rangle$ are the $i$th instantaneous eigenvalue and eigenstate of the full potential, respectively. Here, $n$ is equal to the corresponding quantum number of $|\psi_{R_i}\rangle$ in the full potential at $t = t_m$, which is 3566 for Fig. 3(d) of the main text.

For stage I, assuming that the right well is harmonic, we have $H = \hbar\omega a^\dagger a$ with $\omega = \sqrt{2ea_2/m_e}$. Then, we only need

to consider the harmonic ground state $|0\rangle$ ($|\psi_{R_1}\rangle$) and the excited state $|2\rangle$ ($|\psi_{R_3}\rangle$) in Eq. (S14) since $\partial H/\partial a_2 = z^2$. By substituting the parameter $a_2$ with $\omega$, we find

$$\frac{d\omega}{dt} = -\frac{\epsilon}{\hbar}\frac{(2\hbar\omega)^2}{\sqrt{2\hbar}} = -\epsilon 2\sqrt{2}\omega^2. \tag{S15}$$

This can be integrated to yield

$$\omega(t) = \frac{1}{\epsilon 2\sqrt{2}t + 1/\omega(t_0)} \tag{S16}$$

and the definition of $\omega$ furthermore implies $a_2 = m_e\omega^2/2e$, which thereby gives

$$a_2(t) = \frac{m_e}{2e}\left[\epsilon 2\sqrt{2}t + 1/\omega(t_0)\right]^{-2}. \tag{S17}$$

The parameter $\epsilon$ can also be calculated according to

$$\epsilon = \sqrt{\frac{m_e}{2e}}\frac{\sqrt{1/a_2(t_m)} - \sqrt{1/a_2(t_0)}}{2\sqrt{2}t_m}. \tag{S18}$$

In our simulations, we adopt a unit system with the length unit $\tilde{L} = 1$ $\mu$m, energy unit $\tilde{E} = \hbar^2/(m_e\tilde{L}^2)$, time unit $\tilde{T} = m_e\tilde{L}^2/\hbar$, and charge unit $\tilde{Q} = e$, which yields 1 $\mu$m $= 1$, $\hbar = 1$, $m_e = 1$ and $e = 1$ in such unit system. Equations (S17) and (S18) thereby become

$$a_2(t) = \left[4\epsilon t + \sqrt{1/a_2(t_0)}\right]^{-2}, \tag{S19}$$

$$\epsilon = \frac{\sqrt{1/a_2(t_m)} - \sqrt{1/a_2(t_0)}}{4t_m}. \tag{S20}$$

For stage II, following Ref. [8], we integrate Eq. (S14) from $t = t_m$ to $t = t_f$ using separation of variables, viz.,

$$\epsilon t_2 = -\hbar \int_{a_2(t_m)}^{a_2(t_f)} \left\{\min_{i\neq n}\left|\frac{[E_n(a_2) - E_i(a_2)]^2}{\langle\psi_n(a_2)|\frac{\partial H}{\partial a_2}|\psi_i(a_2)\rangle}\right|\right\}^{-1} da_2, \tag{S21}$$

where $t_2 = t_f - t_m$. We then define $\tilde{\epsilon} = t_2\epsilon$, which is, in fact, independent of $t_2$ or $\epsilon$, based on Eq. (S21). Thus, we solve Eq. (S14) once setting $\epsilon = 1$, in order to obtain $\tilde{\epsilon}$ and $\tilde{a}_2(t)$, and then scale the result for any $\epsilon$, according to $a_2(t) = \tilde{a}_2(\epsilon t)$ and $t_2 = \tilde{\epsilon}/\epsilon$. During stage II, $a_2$ can be replaced by $a_{2,\text{tiny}}$, since $a_{2,\text{reg}}$ is a constant during this stage. For $a_2(t_m)$ from Fig. 3(c) of the main text, we numerically calculate the left side of Eq. (S14) for 15 equally spaced $a_{2,\text{tiny}}$ with $i$ ranging from $n - 30$ to $n + 30$. This range of $i$ is properly chosen since the right-hand side of Eq. (S14) only involves the minimum. With the discreet data of $da_{2,\text{tiny}}(t)/dt$, we numerically obtain the trajectory of $\tilde{a}_{2,\text{tiny}}(t)$ for $\epsilon = 1$ using `DifferentialEquations.jl` [9]. Combined with the scaling method mentioned above, we obtain $a_{2,\text{tiny}}(t)$ for arbitrary $\epsilon$.

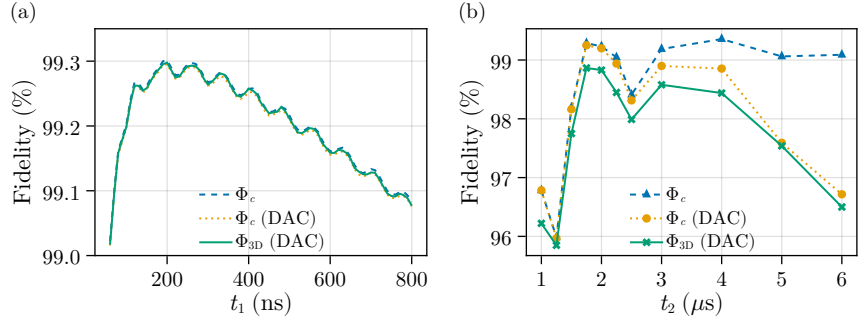

FIG. S7. (a) Fidelity between the ideal eigenstate and the prepared state after stage I as a function of the stage-I evolution time ($t_1 = t_m - t_0$). $\Phi_c$ means that the simulation is performed using the 1D effective potential, DAC means that we consider the effect of DAC during the simulation, whereas $\Phi_{3D}$ means that the simulation is done using the 3D potential. For $\Phi_c$, the fidelity is defined by $F = |\langle \psi_n | \psi \rangle|^2$, whereas, for $\Phi_{3D}$, it is defined by Eq. (6) of the main text ($n = 3566$). (b) Fidelity between the ideal eigenstate and the prepared state after stage I and stage II with the optimal stage-I evolution time 199.4 ns as a function of the stage-II evolution time ($t_2 = t_f - t_m$).

### D. Effect of DAC on system initialization

Here, we discuss the effect of DAC on the system initialization. As mentioned in Sec. I F, the changes of the voltages supplied to the electrodes are always discrete, which leads to a deviation of the generated potentials from our expectation. In order to characterize the impact of this on the system initialization, we simulate the evolution under $\Phi_c$ with and without the DAC, as shown in Fig. S7. It can be clearly seen that the DAC effect is negligible in stage I. However, in stage II, it diminished the performance of the initialization for a long time evolution. We have also simulated the evolution under $\Phi_{3D}$ with the effect of DAC, and the simulation result is nearly identical to the one under $\Phi_c$ with DAC, as shown in Fig. S7. Therefore, the conclusion drawn above can also be applied to real 3D potentials.

Since the DAC influences the generated potential, the data for Fig. 3(c) of the main text is carefully generated. The left and central panels of Fig. 3(c) are obtained with $a_2 = a_{2,\text{reg}} + a_{2,\text{tiny}}(t_m)$. Ignoring the effect of DAC on $a_{2,\text{reg}}$ and $a_{2,\text{tiny}}(t_m)$, smooth lines in the space of eigenenergies-$a_2$ can be produced. The right panel of Fig. 3(c) is obtained with $a_2 = a_{2,\text{reg}}(t_m) + a_{2,\text{tiny}}$, where the DAC effect on $a_{2,\text{tiny}}$ is ignored to give smooth lines and the DAC effect on $a_{2,\text{reg}}(t_m)$ is considered to make the two-level system consistent with the real situation.

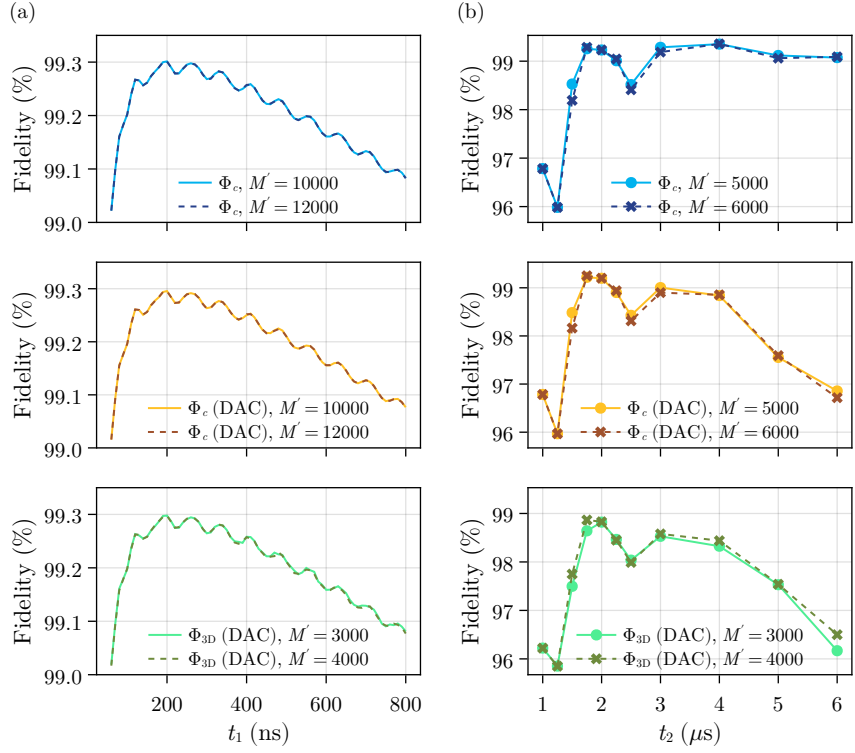

FIG. S8. (a) Fidelity between the ideal eigenstate and the prepared state after stage I as a function of $t_1$ with different $M'$. (b) Fidelity between the ideal eigenstate and the prepared state after stage I and stage II as a function of $t_2$ with different $M'$. Since the evolution time of stage II is larger than that of stage I, we lower $M'$ from 10000 (without DAC) and 12000 (with DAC) for $\Phi_c$ to 5000 and 6000 to speed up the simulation. For $\Phi_{3D}$ with DAC $M' = 3000$ and $M' = 4000$ is used.

### E. Numerical details on the state evolution during the system initialization

The state evolution during the system initialization is simulated in real space using the Trotter expansion [10]

$$e^{\delta(A+B)} = e^{\frac{\delta}{2}A} e^{\delta B} e^{\frac{\delta}{2}A} + O(\delta^3). \tag{S22}$$

In our case, we substitute $A$ with the potential $V$ and $B$ with the kinetic energy $T$ in every step. Since $\exp(\lambda V/2)$ is diagonal in the position basis whereas $\exp(\lambda T)$ is diagonal in the momentum basis, we speed up the calculation by utilizing the fast Fourier transformation to convert the state between the two basis according to

$$e^{\frac{\delta}{2}V} e^{\delta T} e^{\frac{\delta}{2}V} |\psi(t)\rangle_x = e^{\frac{\delta}{2}V_x} \mathcal{T}_{xp}[e^{\delta T_p} \overbrace{\mathcal{T}_{px}[e^{\frac{\delta}{2}V_x} \underbrace{|\psi(t)\rangle_x}_{\text{momentum basis}}]}^{\text{position basis}}], \tag{S23}$$

where $\mathcal{T}_{px}[\cdot]$ and $\mathcal{T}_{xp}[\cdot]$ respectively represent the fast Fourier transform and its inverse, converting the state from position to momentum basis and vice versa. Furthermore, $V_x$ means that we expand $V$ in the position basis, $T_p$ means that we expand $T$ in

the momentum basis, and $|\psi(t)\rangle_x$ denotes that we expand the state in the position basis. The position basis we use has the three ranges $-60.5\,\mu\text{m} < z < 12.5\,\mu\text{m}$, $-0.33\,\mu\text{m} < x < 0.33\,\mu\text{m}$, and $-0.33\,\mu\text{m} < y < 0.33\,\mu\text{m}$, with the sample-point numbers $N_z = 8192$, $N_x = 16$ and $N_y = 16$. These ranges are chosen such that the population outside of them is negligible and the sample-point numbers are chosen so as to ensure the convergence of the relevant eigenstates. As for the number of steps $M$, we choose a dynamical way to control the total error. If we simply assume the error is accumulated during the evolution, the total error during the evolution is $O(M\Delta t^3) = O(t\Delta t^2)$. We define

$$\Delta t = \frac{1\,\text{ns}}{M'\sqrt{\frac{t}{100\,\text{ns}}}}, \tag{S24}$$

where $M'$ represents the number of step per nanosecond for $t = 100$ ns, to give $t\Delta t^2 = (100/M'^2)$ ns$^3$. Thereby, we can roughly control the total error by directly controlling $M'$. The lines in Fig. 3(e) of the main text is obtained using $\Phi_{3D}$ considering DAC and $M' = 4000$. The convergence in Fig. S7 is shown in Fig. S8. All calculations here are done with GPU acceleration via CUDA.jl [11].

## III. TRANSITION FREQUENCY AND EDM MAGNITUDE UNDER DIFFERENT $d$

In this section, we first give details on how Fig. 2 of the main text is obtained. Then, we analyze the scaling of the EDM magnitude with the transition frequency, i.e., the $\mu$-$\omega$ scaling for states generated using our design, Rydberg states, and harmonic-oscillator Fock states. Lastly, we discuss other effects of increasing the potential parameter $d$.

### A. Details on obtaining Fig. 2 of the main text

Figure 2 of the main text is obtained using the ideal 1D potentials of the form

$$a_2(t_f)z^2 + a_3 z^3 + a_4 z^4, \tag{S25}$$

where $a_2(t_f)$ is the value of $a_2$ at the end of stage II. Our method of determining $a_2(t_f)$ ensures that the requirements mentioned in Sec. II B are satisfied for different values of $d$. Here, we introduce a new parameter $a_{2\text{pre}}(d)$, which leads to a constant ratio $c$ between the barrier energy of the potential $a_{2\text{pre}}z^2 + a_3 z^3 + a_4 z^4$ measured relative to the minimum of the right well and the ground-state energy of the potential $a_{2\text{pre}}z^2$. Then, we search for a GRAP near $a_{2\text{pre}}$ and denote the eigenstates involved in the anticrossing at the GRAP by $|\psi_n\rangle$ and $|\psi_{n+1}\rangle$. The value of $a_2(t_f)$ is chosen to be smaller than the GRAP and give a local maximum of $\min(E_{n+1}-E_n, E_n-E_{n-1})$, which ensures $a_2(t_f)$ is away from any GRAP and thereby makes the eigenstates $|\psi_n\rangle$ and $|\psi_{n-1}\rangle$ almost completely confined to the left well, little affected by the small value of $a_2(t_f)$. The two-level system is

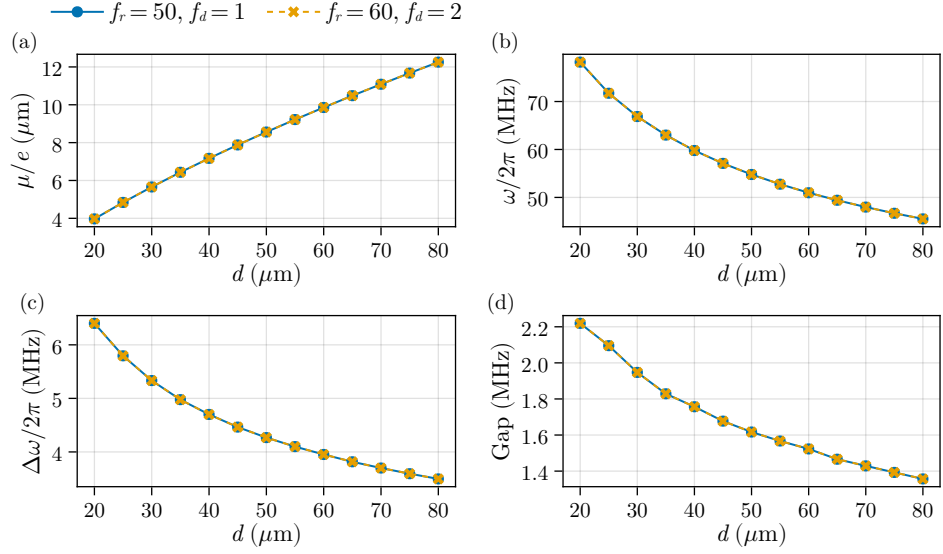

FIG. S9. (a) Magnitude of the EDM as a function of $d$ with different values of $f_r$ and $f_d$. (b) Transition frequency as a function of $d$ with different values of $f_r$ and $f_d$. (c) Effective detuning as a function of $d$ with different values of $f_r$ and $f_d$. (d) Anticrossing gap of the GRAP near $a_{2pre}$ as a function of $d$ with different values of $f_r$ and $f_d$.

composed of $|\psi_n\rangle$ and $|\psi_{n-1}\rangle$. Figure 2 of the main text is obtained using the ratio $c = 2.3$, which ensures that the anticrossing gap of the GRAP near $a_{2\text{pre}}$ lies between $(2\pi)2.3$ MHz and $(2\pi)1.3$ MHz for $20~\mu\text{m} < d < 80~\mu\text{m}$.

We then discuss the numerical details on obtaining Fig. 2 of the main text. Empirically, the range of the position basis for different $d$ can be decided by

$$\Phi(z_{\min}) = \Phi(z_{\max}) = f_r \hbar \omega_z / e \tag{S26}$$

where $\Phi(z) = a_3 z^3 + a_4 z^4$, $\omega_z = (2\pi)300$ MHz, and $f_r$ is a parameter controlling the position range. Equation (S26) ensures that the wavefunctions of the eigenstates with eigenenergies much smaller than $f_r \hbar \omega_z$ are contained in our range. The sample points of the position basis is required to be large enough to yield converging results. Empirically, the sample points of the position basis for different $d$ can be decided by

$$N_z = 1024 \times \left\lfloor \left(\frac{d}{20}\right)^{2.2} + f_d \right\rfloor, \tag{S27}$$

where $\lfloor \cdot \rfloor$ is the floor function returning the largest integer smaller than the given number, and $f_d$ is a parameter controlling the sample points. The value 2.2 in Eq. (S27) is an empirical choice to obtain appropriate $N_z$ for different $d$. Equation (S27) takes into account that the quantum number of the eigenstate we want to prepare increases with increasing $d$. We can change $f_r$ and $f_d$ to test the convergence of the results, as shown in Fig. S9. Figure 2 of the main text is obtained with $f_r = 60$ and $f_d = 2$.

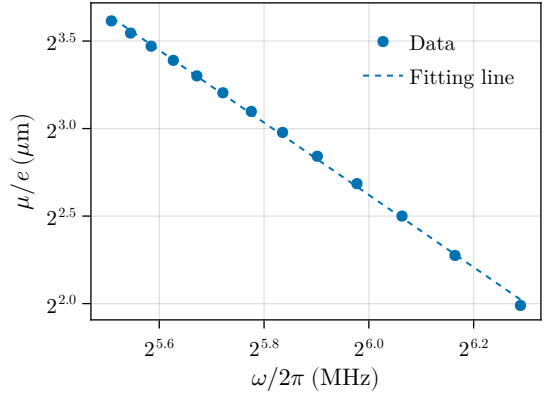

FIG. S10. Log-log plot of the numerical results (solid circles) and the corresponding fit (dashed line) of the relation between EDM and resonance frequency.

### B. Analysis of the $\mu$-$\omega$ scaling

Changing Fig. 2(b) of the main text into double-logarithmic plot results in scatters shown in Fig. S10 that suggests a linear fit. This is described by

$$\log_2(\mu/e) = k \log_2(\omega/2\pi) + b, \tag{S28}$$

where $k$ and $b$ are coefficients that need to be determined by the fit. The 95% confidence interval of $k$ is $[-2.109, -2.022]$ and that of $b$ is $[14.760, 15.269]$. This means we can, in good confidence, conclude $\mu \propto \omega^{-2}$, as mentioned in the main text. With these results, we obtain the blue line in Fig. 2(b) of the main text.

As for Rydberg states, we give a simple estimation on the scaling using the Bohr model and assuming the dipole between neighboring Rydberg states to be proportional to $n^2$ [12]. Within the Bohr model, we have $\omega_n \propto n^{-3}$, where $\omega_n$ represents the transition frequency between the two Rydberg states with principle quantum numbers $n$ and $n-1$. Combined with $\mu \propto n^2$, we thereby find $\mu \propto \omega^{-2/3}$ for the neighboring Rydberg states.

As for harmonic-oscillator Fock states, we directly calculate the magnitude of the EDM between the ground state and the single-phonon state, which can be expressed as $\langle 1|ex|0\rangle$ with $x = \sqrt{\hbar/2m\omega}(a^\dagger + a)$ representing the position operator. This yields the scaling $\langle 1|ex|0\rangle \propto \omega^{-1/2}$.

### C. Other effects of increasing $d$

As $d$ increases, both $a_3$ and $a_4$ decrease according to Eq. (2) of the main text, leading to a reduction of the nonlinear electric-field strength. Therefore, increasing $d$ is beneficial for the realization of the potential $a_3 z^3 + a_4 z^4$, and enables an increase in ion-surface distance, which reduces the electric-field noise. However, there are some constraints on the increase of $d$. A smaller transition frequency induced by a larger $d$ necessitates smaller gaps of the GRAPs near $a_2(t_m)$ to let the state easily cross energy levels at the GRAPs. This, in turn, results in a slower evolution during stage II and a deteriorated system-initialization fidelity. A more fundamental constraint is that $d$ must be considerably smaller than the wavelength of the RF field [e.g., 6 mm for a 50 GHz RF field], which insures that the wave character of the RF field does not affect the electron motion. Based on our numerical calculations for $d = 80$ $\mu$m, the transition frequency is around $(2\pi)45$ MHz, and the gap of the GRAP near $a_2(t_f)$ or $a_2(t_m)$ is roughly $(2\pi)1.3$ MHz. These results suggest the feasibility of a design with $d = 80$ $\mu$m, where the magnitude of EDM exceeds 12 $e\mu$m.

## IV. DETAILS ON STATE READOUT, COHERENT CONTROL AND QUANTUM SENSING

In this section, we first give detailed derivations of $H_B^{(\text{int})}$, $H_E^{(\text{int})}$, and the effective detuning. Subsequently, we discuss the dynamics of our system in a noisy environment. Finally, we give details on the numerical simulations of the state readout and electric-field sensing.

### A. Derivations of the Hamiltonians and the effective detuning

In the interaction picture with respect to $H_z$, Eq. (7) of the main text can be rewritten as

$$\begin{aligned} H_B^{(\text{int})} &= -\mu_\text{B} b_y s_y \cos\left(\omega_B t + \phi_B\right) \sum_{i,j} \exp[i\left(\omega_i - \omega_j\right) t] z_{ij} \left|\psi_i\right\rangle\!\left\langle\psi_j\right| \\ &= -\frac{1}{2}\mu_\text{B} b_y s_y \{\exp[i(\omega_B t + \phi_B)] + \exp[-i(\omega_B t + \phi_B)]\} \sum_{i,j} \exp(i\omega_{ij} t) z_{ij} \left|\psi_i\right\rangle\!\left\langle\psi_j\right|, \end{aligned} \quad \text{(S29)}$$

where $b_y = \partial B_y/\partial z$ and $s_y$ is the $y$ Pauli matrix of the spin degree of freedom. Additionally, we have the eigenvalue equation $H_z \left|\psi_i\right\rangle = \hbar\omega_i \left|\psi_i\right\rangle$, transition frequencies $\omega_{ij} = \omega_i - \omega_j$, and matrix elements $z_{ij} = \left\langle\psi_i|z|\psi_j\right\rangle$. If the two-level system is composed of the $n$th and $n'$th eigenstates and the system is initialized into one of these two eigenstates, we can ignore the terms in Eq. (S29) that satisfy $i \neq n, n'$ and $j \neq n, n'$. After that, considering $\omega_B = \omega_{nn'} > 0$ and applying a rotating-wave

approximation to further ignore fast-oscillating terms, we arrive at

$$\begin{aligned}H_B^{(\text{int})} &= -\frac{1}{2}\mu_B b_y s_y[\exp(i\phi_B)z_{n'n}|\psi_{n'}\rangle\langle\psi_n| + \exp(-i\phi_B)z_{nn'}|\psi_n\rangle\langle\psi'_n|]\\ &= -\frac{1}{2}\mu_B b_y |z_{nn'}| s_y[\exp(i\phi)|\psi_{n'}\rangle\langle\psi_n| + \exp(-i\phi)|\psi_n\rangle\langle\psi'_n|]\\ &= -\frac{1}{2}\mu_B b_y |z_{nn'}| s_y[\cos(\phi)\,\sigma_x + \sin(\phi)\,\sigma_y]\\ &= -\frac{1}{2}\hbar g s_y \sigma_\phi,\end{aligned} \quad (S30)$$

where we have defined $\sigma_\phi = \cos(\phi)\sigma_x + \sin(\phi)\sigma_y$, with the Pauli operators $\sigma_x$ and $\sigma_y$ of the two-level system and the $+1$ eigenstate $|\psi_n\rangle$ of $\sigma_z$. Furthermore, we use the phase $\phi = \phi_B - \phi_z$, with $\phi_z = \arg(z_{nn'})$, and coupling strength $g = \mu_B b_y |z_{nn'}|/\hbar$. Assuming that the spin is initialized into the state $|\uparrow\rangle$ (the $+1$ eigenstate of $s_z$) and the motional state has the form $\sin(\phi_m)|\uparrow_\phi\rangle + \cos(\phi_m)|\downarrow_\phi\rangle$, after the time $t = \pi/2g$, the full quantum state has evolved into

$$\sin(\phi_m)|\uparrow_\phi\rangle|-\rangle + \cos(\phi_m)|\downarrow_\phi\rangle|+\rangle, \quad (S31)$$

where $|\uparrow_\phi\rangle$ ($|\downarrow_\phi\rangle$) represents the $+1$ ($-1$) eigenstate of $\sigma_\phi$ and $|\pm\rangle$ likewise represents the $\pm 1$ eigenstate of $s_x$.

As for coherent control, we consider an oscillating electric field along the axial direction with the amplitude $\mathcal{E}_z$ and a frequency $\omega_E = \omega_{nn'} > 0$ that is resonant with the two-level-system transition. The Hamiltonian can be written as

$$H_E = H_z - ez\mathcal{E}_z \cos(\omega_E t + \phi_E). \quad (S32)$$

In the interaction picture with respect to $H_z$, we find

$$H_E^{(\text{int})} = -e\mathcal{E}_z \cos(\omega_E t + \phi_E)\sum_{i,j}\exp(i\omega_{ij}t)z_{ij}|\psi_i\rangle\langle\psi_j| \quad (S33)$$

and under a rotating-wave approximation this becomes

$$H_E^{(\text{int})} = \frac{1}{2}|\mu_{nn'}|\mathcal{E}_z \sigma_{m-\phi'} = \frac{1}{2}\hbar\Omega_R \sigma_{m-\phi'}, \quad (S34)$$

with $\mu_{nn'} = e\langle\psi_n|z|\psi_{n'}\rangle$, $\Omega_R = |\mu_{nn'}|\mathcal{E}_z/\hbar$, $\phi' = \phi_E - \phi_z$, and $\phi_z = \arg(z_{nn'})$.

As for quantum sensing, one finds exactly the same Hamiltonian as the one for coherent control. In the interaction picture, described by Eq. (S34), the state rotates around an axis lying in the $x$-$y$ plane of the Bloch sphere with an angle $\phi'$ to the $x$-axis. The probability to find the system in the $+1$ eigenstate of $\sigma_\phi$ ($\phi = \phi_E - \phi_z - \pi/2$) is given by $p = [1 + \sin(\Omega_R t)]/2$. For a weak electric field, we only need to consider the susceptibility to the electric field strength at $\mathcal{E}_z = 0$, namely

$$\left.\frac{\partial p}{\partial \mathcal{E}_z}\right|_{\mathcal{E}_z=0} = |\mu_{nn'}|t/2\hbar. \quad (S35)$$

If we take the influence of the electric-field noise near the surface into account, the susceptibility follows a decay model [13, 14]

$$\left.\frac{\partial p}{\partial \mathcal{E}_z}\right|_{\mathcal{E}_z=0} = |\mu_{nn'}|\exp(-\Gamma t)t/2\hbar, \tag{S36}$$

where $\Gamma$ is the effective decay rate associated with the transitions between eigenstates induced by the electric-field noise near the surface.

All the above discussions rely on the rotating-wave approximation. Unlike usual two-level systems, in our case, it is not sufficient for $\Omega_R$, $g$, $|\mu_{nn} - \mu_{n'n'}|\mathcal{E}_z/\hbar$, and $\mu_B b_y |z_{nn} - z_{n'n'}|/\hbar$ to be much smaller than $\omega_{nn'}$. However, additionally, it is also necessary that $|\mu_{ij}|\mathcal{E}_z/\hbar$ and $\mu_B b_y |z_{ij}|/\hbar$ are much smaller than $||\omega_{ij}| - |\omega_{nn'}||$ and $||\omega_{ij}| + |\omega_{nn'}||$, for $i \in (n, n'), j \notin (n, n')$, which prevents the quantum state from leaking to other energy levels. Since $||\omega_{ij}| - |\omega_{nn'}|| < ||\omega_{ij}| + |\omega_{nn'}||$, defining $\Delta\omega_{ij,nn'} = ||\omega_{ij}| - |\omega_{nn'}||$, this requirement can be written as

$$\Omega_R, g \ll \frac{|\mu_{nn'}|}{|\mu_{ij}|}\Delta\omega_{ij,nn'}, \text{ for } i \in (n, n'), j \notin (n, n'). \tag{S37}$$

Then, defining the effective detuning $\Delta\omega_{nn'}$ according to Eq. (10) of the main text, we can simplify this requirement to read

$$\Omega_R, g \ll \Delta\omega_{nn'}. \tag{S38}$$

### B. Master equation and transition rates

The dynamics of our systems in a noisy environment can be described by a master equation of the Lindblad form, which reads

$$\dot\rho = -\frac{i}{\hbar}[H, \rho] + \sum_{ij}\Gamma_{ij}\left(J_{ij}\rho J_{ij}^\dagger - \frac{1}{2}J_{ij}^\dagger J_{ij}\rho - \frac{1}{2}\rho J_{ij}^\dagger J_{ij}\right), \tag{S39}$$

where $J_{ij}$ are the jump operators $|\psi_j\rangle\langle\psi_i|$ and $\Gamma_{ij}$ the transition rate from $\psi_i$ to $\psi_j$ ($i \neq j$).

In the following, we present a derivation of the transition rates induced by the electric-field noise near the surface following Ref. [15]. Considering the axial direction ($z$ direction), the Hamiltonian with a fluctuating potential $\Phi(t, \mathbf{r})$ can be written as

$$H(t) = H_z + e\Phi(t, z). \tag{S40}$$

If we consider the effect of the fluctuating potential $\Phi(t, z)$ only to first order, the Hamiltonian can be further simplified to

$$H(t) = H_z - eE_z(t)z, \tag{S41}$$

where $E_z = -\partial\Phi/\partial z$. Using Eq. (S41) and first-order perturbation theory, we can calculate the transition rate from state $|\psi_i\rangle$ to state $|\psi_j\rangle$, $\Gamma_{ij}$. The state of the perturbed system can be expanded in the eigenstates of the unperturbed system according

to $|\psi(t)\rangle = \sum_n a_n(t)e^{-i\omega_n t}|\psi_n\rangle$, where $a_n(t)$ is caused by the perturbation Hamiltonian $H' = -eE_z(t)z$. The Schrödinger equation reads

$$\left(H_z + H' - i\hbar\frac{\partial}{\partial t}\right)\sum_n a_n(t)|\psi_n\rangle e^{-i\omega_n t} = 0 \tag{S42}$$

and yields

$$i\hbar\frac{da_j(t)}{dt} = \sum_n \langle\psi_j|H'|\psi_n\rangle a_n(t)e^{i(\omega_j-\omega_n)t}. \tag{S43}$$

For the initial state $|\psi_i\rangle$, if the perturbation is weak, we assume $a_n(0) = \delta_{n,i}$. From Eq. (S43) we then obtain

$$\frac{da_j(t)}{dt} = \frac{1}{i\hbar}\langle\psi_j|H'|\psi_i\rangle e^{i(\omega_j-\omega_i)t}, \tag{S44}$$

which can be integrated according to

$$a_j(t) = \frac{1}{i\hbar}\int_0^t \langle\psi_j|H'|\psi_i\rangle e^{i(\omega_j-\omega_i)t}dt. \tag{S45}$$

If we define the transition probability $P_{ij}(t) = |a_j(t)|^2$ and substitute $H'$ with $-eE_z(t)z$, we obtain

$$P_{ij}(t) = \frac{|\mu_{ij}|^2}{\hbar^2}\left|\int_0^t E_z(t)e^{-i\omega_{ij}t}dt\right|^2. \tag{S46}$$

We then define the autocorrelation function of the electric field $E_z(t)$ as

$$R_{EE}(t_1, t_2) = \langle E_z(t_1)E_z(t_2)\rangle. \tag{S47}$$

For a wide-sense stationary noise process, $R_{EE}(t_1, t_2)$ only relates to $\tau = t_2 - t_1$ [15], and we can have

$$R_{EE}(\tau) = \frac{1}{t}\int_0^t E_z(t_1)E_z(t_1+\tau)dt_1. \tag{S48}$$

In terms of this autocorrelation function, the transition probability can be cast into the form

$$\begin{aligned}P_{ij}(t) &= \frac{|\mu_{ij}|^2}{\hbar^2}\int_0^t E_z(t_1)e^{i\omega_{ij}t_1}dt_1\int_0^t E_z(t_2)e^{-i\omega_{ij}t_2}dt_2 \\ &= \frac{|\mu_{ij}|^2}{\hbar^2}\int_0^t E_z(t_1)e^{i\omega_{ij}t_1}dt_1\int_{-t_1}^{t-t_1} E_z(t_1+\tau)e^{-i\omega_{ij}(t_1+\tau)}d\tau \\ &= \frac{|\mu_{ij}|^2}{\hbar^2}\int_0^t dt_1\int_{-\infty}^{\infty}d\tau e^{-i\omega_{ij}\tau}E_z(t_1)E_z(t_1+\tau) \\ &= t\frac{|\mu_{ij}|^2}{\hbar^2}\int_{-\infty}^{\infty}e^{-i\omega_{ij}\tau}R_{EE}(\tau)d\tau.\end{aligned} \tag{S49}$$

Here, we assume that $t$ is short compared to the timescale of the population evolution, but large compared to the correlation time of $E_z(t)$, such that the integration range of $\tau$ can formally be extended to $\pm\infty$[16]. We now define the spectral density of the electric-field noise, given by

$$S_E(\omega_{ij}) = 2\int_{-\infty}^{\infty}e^{-i\omega_{ij}\tau}R_{EE}(\tau)d\tau, \tag{S50}$$

and obtain the transition rate

$$\Gamma_{ij} = \frac{P_{ij}(t)}{t} = \frac{1}{2}\frac{|\mu_{ij}|^2}{\hbar^2}S_E(\omega_{ij}). \tag{S51}$$

The electric-field noise near the surface has multiple sources, e.g., blackbody radiation, Johnson-Nyquist noise, and patch potentials, to mention only a few [15]. However, despite extensive research, the picture of the underlying mechanisms is still incomplete [17]. Therefore, instead of calculating the electric-field-noise spectral density from first principles, we infer it by extrapolating the experimental results obtained in ion traps. Assuming the electric-field noise spectral density follows a power law $S_E(\omega) \propto 1/\omega^{1.3}$ [18] and $S_E[(2\pi)1 \text{ MHz}] \approx 10^{-12}$ (V/m)$^2$/Hz in a 4-K environment with a particle-surface distance of 30 $\mu$m [3, 15], we can extrapolate the spectral density to the frequency range of our system. Then, with Eq. (S51), we can obtain the induced transition rates between eigenstates.

Additionally, since our two-level system is composed of eigenstates with large quantum numbers, we can also calculate the rate of spontaneous transitions from our two-level system to other lower-lying eigenstates, which have there origin in the coupling to the free electromagnetic field. This spontaneous transition rate from the eigenstate $|\psi_i\rangle$ to $|\psi_j\rangle$ ($i > j$) can be written as

$$\Gamma_{ij}^s = \frac{|\mu_{ij}|^2\omega_{ij}^3}{3\pi\epsilon_0\hbar c^3}, \tag{S52}$$

where $c$ is the speed of light and $\epsilon_0$ is the vacuum permittivity [19]. Based on our calculation, the maximum spontaneous transition rate is below $3 \times 10^{-3}$ Hz for our system described in Fig. 3 of the main text. This means that these spontaneous transitions are negligible compared to the ones induced by electric-field noise, whose magnitude inferred from the extrapolation is typically in the hundreds-of-kHz range. Therefore, we can ignore the spontaneous transitions in our system.

### C. Numerical details on state readout and sensing

Numerically, with Eq. (S51) and the electric-field noise density, the transition rate between the two-level system described in Fig. 3 of the main text is $\Gamma_{3566,3565} = 289.6$ kHz. The sum of transition rates from the two-level system to other eigenstates except those with quantum number from 3561 to 3575 is less than 7 kHz. Therefore, we can neglect eigenstates with quantum number larger than 3575 or smaller than 3561 and do the simulations of state readout and electric-field sensing with a motional subspace formed by eigenstates with quantum number from 3561 to 3575. Using `QuantumOptics.jl` [20], we accomplish the simulation via `ode45` (`DP5`) adaptive method with the step-size control options $abstol = 1.0 \times 10^{-9}$ and $reltol = 1.0 \times 10^{-7}$. These options ensure that the local error, $err$, satisfies $err < abstol + max(uprev, u) * reltol$, where $u$ and $uprev$ represent the vector elements of the current and previous steps, respectively [9]. Figure S11 shows the convergence of the simulation results of averaged fidelity and $\partial p/\partial \mathcal{E}_z|_{\mathcal{E}_z=0}$.

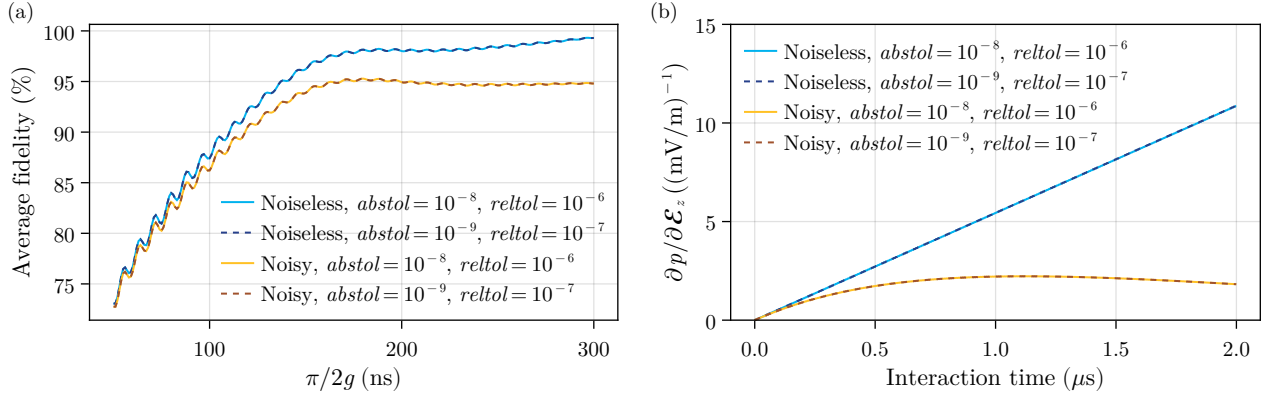

FIG. S11. (a) Average fidelity as a function of the transfer time $\pi/2g$ with different step-size control options. (b) $\partial p/\partial \mathcal{E}_z$ as a function of the interaction time at $\mathcal{E}_z = 0$ with different step-size control options.

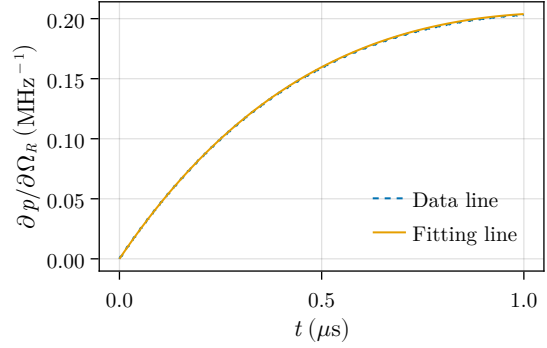

FIG. S12. Fitting $\partial p/\partial \Omega_R|_{\Omega_R=0}$ with the decay model.

The average fidelity is obtained using $F_{\text{avg}} = (F_+ + F_-)/2$, where $F_\pm$ represents the fidelity between the final spin state and the ideal state $|\pm\rangle$ with the motional state prepared in the $-1$ ($+1$) eigenstate of $\sigma_\phi$. To obtain $\partial p/\partial \mathcal{E}_z|_{\mathcal{E}_z=0}$, we calculate $\partial p/\partial \Omega_R|_{\Omega_R=0}$ by finite differences [21] and then use the chain rule $\partial p/\partial \mathcal{E}_z = (\partial p/\partial \Omega_R)(\partial \Omega_R/\partial \mathcal{E}_z)$ with $\partial \Omega_R/\partial \mathcal{E}_z = |\mu_{n_1 n_2}|/\hbar$. Considering a noisy environment, we then find

$$\left.\frac{\partial p}{\partial \Omega_R}\right|_{\Omega_R=0} = \frac{1}{2}\exp(-\Gamma t)t, \tag{S53}$$

where $\Gamma$ is the effective decay rate. Fitting the dashed blue line in Fig. 4(b) of the main text with model described by Eq. (S53), we find $\Gamma = 0.8972$ MHz with a 95% confidence interval $[0.89711, 0.89734]$ MHz. The data and corresponding fit are shown in Fig. S12.



\end{document}